\pdfoutput=1
\documentclass[sigconf, nonacm,dvipsnames]{acmart}
\usepackage{booktabs}
\usepackage{multirow}
\usepackage{tikz}
\usepackage{amsmath}
\usepackage{siunitx}
\usepackage[abbreviations]{foreign}
\usepackage{xspace}
\usepackage[inline]{enumitem}
\usepackage[algo2e,ruled,lined,boxed,commentsnumbered, noend, linesnumbered, procnumbered]{algorithm2e}
\definecolor{orange}{rgb}{1,0.5,0}
\definecolor{Green}{rgb}{0,0.5,0}
\definecolor{Blue}{rgb}{0,0,1}
\definecolor{OliveGreen}{rgb}{0.33,0.42,0.18}

\usepackage{framed}
\definecolor{takeawaycolor}{RGB}{232, 244, 232}
\newenvironment{takeaway}{\colorlet{shadecolor}{takeawaycolor}\begin{shaded}\noindent\textbf{Takeaway.} }{\end{shaded}}
\usepackage{pifont}%
\newcommand{\cmark}{\ding{51}}%
\newcommand{\xmark}{\ding{55}}%
\newcommand{\greencheck}{{\color{OliveGreen}\cmark}}
\newcommand{\redcross}{{\color{red}\xmark}}
\usepackage{ifthen}
\usepackage{acronym}
\acrodef{ANN}{approximate nearest neighbour}
\acrodef{NNS}{nearest neighbour search}
\acrodef{IVF}{inverted file}
\acrodef{LWE}{learning with errors}
\acrodef{LPN}{learning parity with noise}
\acrodef{LHE}{linearly-homomorphic encryption}
\acrodef{FHE}{fully-homomorphic encryption}
\acrodef{BFV}{Brakerski-Fan-Vercauteren}
\acrodef{PIR}{private information retrieval}
\acrodef{DCPE}{distance-comparison-preserving encryption}
\acrodef{SAP}{scale-and-perturb}
\acrodef{TDM}{trapdoored matrix}
\acrodef{EMVP}{encrypted matrix-vector product}
\acrodef{LLM}{large language model}
\acrodef{RAG}{retrieval-augmented generation}
\acrodef{GPU}{graphics processing unit}
\acrodef{CPU}{central processing unit}
\acrodef{DRAM}{dynamic random-access memory}
\acrodef{NUMA}{non-uniform memory access}
\acrodef{SIMD}{single-instruction multiple-data}
\acrodef{SMT}{simultaneous multithreading}
\acrodef{CSV}{comma-separated values}
\acrodef{CDF}{cumulative distribution function}
\acrodef{RNG}{random number generator}
\acrodef{CUDA}{Compute Unified Device Architecture}
\acrodef{AI}{artificial intelligence} %
\newcommand{\plaintext}{\textsc{Plaintext}\xspace}
\newcommand{\sap}{\textsc{SAP}\xspace}
\newcommand{\emvp}{\textsc{EMVP}\xspace}
\newcommand{\bntm}{\textsc{BNTM}\xspace}
\newcommand{\tiptoe}{\textsc{Tiptoe}\xspace}

\newcommand{\msmarco}{MS~MARCO\xspace}
\newcommand{\msmarcosmall}{MS~MARCO-100k\xspace}
\newcommand{\msmarcofull}{MS~MARCO-8.8M\xspace}
\newcommand{\efive}{E5-base-v2\xspace}

\graphicspath{ {figures/} }
\usepackage{tikz}
\usepackage[eulergreek]{sansmath}
\usepackage{pgfplots}
\usepackage{pgfplotstable}
\usepackage{cleveref}
\usepackage{comment}
\crefname{assumption}{assumption}{assumptions}

\pgfplotsset{compat=newest}
\usepgfplotslibrary{external,units,colorbrewer,groupplots,fillbetween,statistics}
\usetikzlibrary{patterns,shapes.misc}

\usepackage[eulergreek]{sansmath}
\usepackage{pgfplots}
\usepackage{pgfplotstable}

\pgfplotsset{compat=newest}
\usepgfplotslibrary{units,colorbrewer,groupplots,fillbetween,statistics}
\usetikzlibrary{arrows.meta,backgrounds,calc,fit,shapes.geometric,positioning,patterns,shapes.misc}

\pgfplotsset{
  width=3.25in, height=2in,
  every axis/.append style={
    semithick,
    ymajorgrids=true,
    label style={font=\sffamily\small},
    title style={font=\sffamily\small},
  },
  every axis plot/.append style={thick},
  legend style={
    draw=none,
    fill=none,
    text opacity=1,
    anchor=south,
    at={(0.5,1)},
    font=\sffamily\small,
  },
  legend cell align=left,
  legend columns=-1,
  xtick align=outside,
  xtick pos=bottom,
  major tick length=1mm,
  tick label style={font=\sansmath\sffamily\small},
  every axis label={font=\sffamily\small},
  label style={font=\sffamily\small},
  {cycle list/Dark2},
  {cycle list/Set1},
  cycle list name=Dark2,
  cycle multiindex* list={
    Dark2\nextlist
    {solid,dashed,dotted,dashdotted,densely dashdotdotted}\nextlist
    mark list*\nextlist
  },
  select coords between index/.style 2 args={
    x filter/.code={
      \ifnum\coordindex<#1\fi
      \ifnum\coordindex>#2\fi
    }
  },
  discard if not/.style 2 args={
    x filter/.code={
      \edef\tempa{\thisrow{#1}}
      \edef\tempb{#2}
      \ifx\tempa\tempb
      \else
        
      \fi
    }
  },
  discard if above/.style 2 args={
    x filter/.code={
      \edef\tempa{\thisrow{#1}}%
      \ifdim\tempa pt>#2pt
      \fi
    }
  },
}

\newcommand{\inputplot}[1]{\includegraphics{#1.pdf}}

\newcommand{\newgroupwidth}[2]%
{\expandafter\xdef\csname groupwidth#1\endcsname{#2}}

\newcounter{groupwidth}
\newsavebox{\groupwidthbox}
\makeatletter
{\edef\groupnumber{#1}%
	\stepcounter{groupwidth}%
	\@ifundefined{groupwidth\thegroupwidth}{\pgfmathsetlengthmacro{\mywidth}{\linewidth/\groupnumber}}%
	{\expandafter\let\expandafter\mywidth\csname groupwidth\thegroupwidth\endcsname}%
	\begin{lrbox}{\groupwidthbox}%
		\tikzset{/pgfplots/width={\mywidth}}%
		\ignorespaces}%
	{\end{lrbox}%
	\usebox\groupwidthbox
	\pgfmathsetlengthmacro{\mywidth}{\mywidth + (\linewidth - \wd\groupwidthbox)/\groupnumber}
	\immediate\write\@auxout{\string\newgroupwidth{\thegroupwidth}{\mywidth}}}
\makeatother

\usepackage{amsmath,amssymb,amsfonts,amsthm}
\usepackage{enumitem}
\usepackage{thm-restate}
\usepackage{bbm}
\theoremstyle{definition}

\theoremstyle{remark}

\allowdisplaybreaks

\makeatletter
\AtBeginDocument{%
	\def\ltx@label#1{\cref@label{#1}}%

	\def\label@in@display@noarg#1{\cref@old@label@in@display{#1}}%

	\def\label@in@mmeasure@noarg#1{%
		\begingroup
		\measuring@false
		\cref@old@label@in@display{#1}%
		\endgroup
	}%
}
\makeatother
 \begin{document}
\title{A Unified Benchmark for Privacy-preserving Vector Search}

\author{Anne-Marie Kermarrec}
\affiliation{
  \institution{EPFL}
  \city{Lausanne}
  \country{Switzerland}
}

\author{Rafael Pires}
\affiliation{
  \institution{EPFL}
  \city{Lausanne}
  \country{Switzerland}
}

\author{Mathis Randl}
\affiliation{
  \institution{EPFL}
  \city{Lausanne}
  \country{Switzerland}
}

\author{Martijn de Vos}
\affiliation{
  \institution{EPFL}
  \city{Lausanne}
  \country{Switzerland}
}

\renewcommand{\shortauthors}{Kermarrec et al.}

\begin{abstract}

Vector search powers semantic search, recommendation systems, and \ac{RAG}. By design, the service answering a query sees both the query embedding and, usually, the corpus against which it is matched. This is a privacy breach for both the user issuing the query and the owner of the corpus. A family of cryptographic schemes (e.g., \sap, \emvp, \bntm, \tiptoe) addresses that leak. However, as each scheme is published and evaluated on its own corpus, threat model, parameter choices, hardware, and metric conventions, the numbers cannot be compared directly.  Consequently, a practitioner asking which one to deploy today has no defensible way to choose. We close that gap with a uniform experimental comparison, including a \plaintext baseline and four cryptographic backends running over the same workload, hardware, and metric definitions. Under that ruler, the schemes spread across a Pareto frontier in privacy, performance, and recall rather than imposing a flat penalty on performance. We find that the performance of \sap matches \plaintext, \emvp delivers cryptographic indistinguishability at a 4$\times$ throughput cost on CPU, \bntm adds malicious-server verifiability at a further 22$\times$ median-latency cost, and \tiptoe hides the cluster choice itself, but incurs a 190$\times$ per-query cost compared to \plaintext. GPU acceleration pays off for \plaintext and \sap but not for \emvp or \bntm.
All our experiment artifacts are publicly available for reproducibility.
\end{abstract}
 
\maketitle
\pagestyle{plain}

\section{Introduction}
\label{sec:intro}

Vector search is a key mechanism underlying modern information retrieval~\cite{karpukhin2020dense,pan2024survey}.
It involves finding the vectors closest to a given query vector among a collection that can hold billions of them~\cite{simhadri:2022:bigann,jegou:2011:pq}.
Each vector in this collection is typically a high-dimensional embedding representing a document, image, or other data item~\cite{reimers2019sentence}.
Vector search underlies recommendation systems~\cite{rajput2023recommender} and code-search tools used daily by developers~\cite{feng2020codebert}, as well as \ac{RAG} pipelines, where a \ac{LLM} grounds its answers in documents fetched by nearest-neighbor vector search~\cite{lewis2020retrieval}.
It is also an important component of commercial retrieval products and vector databases~\cite{openai:2026:filesearch,pinecone:2026:vectordb} and is included in some agentic frameworks to manage the memory of \ac{LLM} agents~\cite{zhong2024memorybank}.
This makes vector search one of the most widely deployed pieces of infrastructure in retrieval systems nowadays.

Vector search is typically performed without any privacy protection: the service performing the vector search sees both the query embedding and the corpus embeddings against which it is matched.
This exposes both the user issuing the query and the corpus owner to privacy violations.
Research has shown that dense embeddings can be partially inverted to \emph{recover substantial portions of the original text} (embedding inversion)~\cite{song2020information,morris:2023:embedinv}, even when the attacker can only query the embedding model as a black box (\ie, without access to its internal weights).
These inversion attacks allow the server to reconstruct parts of the corpus, and to detect which parts of the corpus are typically queried.
This is problematic in high-stake domains such as healthcare, where a server observing query embeddings can learn which symptoms or conditions clinicians and patients are searching for, and even invert corpus embeddings to obtain the underlying patient records themselves~\cite{tsai2026concept}.

\begin{figure*}[t]
    \centering
    \inputplot{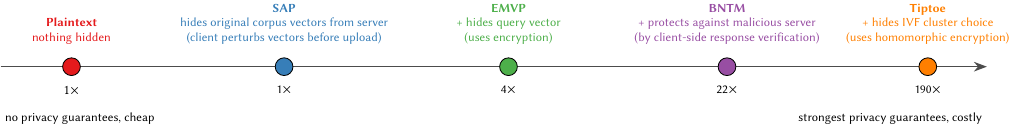}
    \caption{Comparing the five schemes on privacy guarantees and computational overhead. Each marker adds privacy guarantees over the one to its left, at an increase in per-query cost. The cost figures are the median end-to-end latency compared to the unencrypted baseline (\plaintext); \bntm's figure includes response verification. See \Cref{sec:evaluation} for the full experimental setup and results.
    }
    \label{fig:schematic-spectrum}
\end{figure*}

The cryptographic community has answered this concern from substantively different directions.
At the lightest end, \sap~\cite{fuchsbauer:2021:dcpe} scales and perturbs each vector with a small amount of noise before uploading it to the server, in such a way that the server can still compare distances between ciphertexts directly.
However, this still enables the server to learn the geometry of the corpus, \ie which stored vectors lie close to which.
\emvp~\cite{benhamouda:2025:emvp} and \bntm~\cite{braverman:2026:bntm} instead encrypt the vectors themselves and return inner products computed entirely under encryption, hiding both the corpus and the query from the server.
\bntm additionally lets the client verify that the server's answer is correct, a property the other schemes lack.
\tiptoe~\cite{henzinger:2023:tiptoe} goes a step further by layering private information retrieval over the entire pipeline so that the server does not even learn which part of the corpus a query targets.
These schemes therefore span a wide spectrum, from lightweight perturbation to full protection of both the queries and the stored vectors.
This spectrum of privacy guarantees comes with widely different trade-offs in performance and retrieval accuracy.
\Cref{fig:schematic-spectrum} shows the performance cost of each privacy level.

Despite the different options for privacy-preserving vector search, no published study compares them on common ground.
Each scheme is evaluated in its own paper, on its own corpus, with its own embedding model, hardware, and metric conventions that do not always refer to the same definition.
As a result, a practitioner asking today which scheme to deploy has no defensible way to choose because no one has analyzed them under a common ruler.

This work presents a single, uniform comparison of privacy-preserving vector search schemes.
We implement the baseline without any privacy schemes (\plaintext) alongside all four privacy-preserving schemes in one common software harness, run every scheme over the same index, the same workload (the \msmarco text corpus), and the same hardware, and measure all of them using the same metrics.
Under a single benchmark, the differences between the schemes come out more clearly than any individual paper shows.
The results debunk the belief, inherited from early fully-homomorphic prototypes, that encrypted vector search is unaffordable: even with corpus and query fully encrypted, \emvp answers 12 queries per second at recall 0.9 on the full \msmarco corpus with 8.8\,M passages, on one commodity server.
The lightest-weight scheme (\sap) runs at practically the same speed as the unencrypted baseline at recall \(0.9\), on both CPU and \ac{GPU}.
Adopting \sap therefore costs almost nothing in speed, but it cannot guarantee privacy of the corpus itself, since the server can still tell which stored items are similar to one another.
\emvp and \bntm fully encrypt the data but are considerably slower: \emvp is \mbox{4$\times$} slower than the \plaintext baseline.
\bntm, which adds the ability to detect a cheating server, sits further below that (22$\times$ slower than \plaintext).
The strongest scheme, \tiptoe, also hides which part of the corpus a query touches, but is a further 9$\times$ slower than \bntm, roughly two orders of magnitude below \plaintext, on a corpus subset with 100\,k passages.
We also find that the amount of data sent over the network and the amount of data processed in memory rank these schemes differently: \bntm sends very little over the network but reads a large amount from memory internally, and it is this internal memory cost that predicts how well a scheme benefits from a \ac{GPU}.
We find that \ac{GPU} acceleration helps \plaintext and \sap, but not \bntm and \emvp.

This paper makes the following three contributions:
\begin{itemize}
    \item We build a single evaluation harness for privacy-preserving vector search, which includes one corpus, one index configuration, one threat model, and one set of metric definitions that isolate the cryptographic primitive from the rest of the pipeline. This is what makes the four privacy-preserving schemes and the unencrypted baseline comparable to each other  (\Cref{sec:methodology}).
    \item We benchmark all five schemes across recall, throughput, latency, and communication cost. We also measure the dimensions that decide whether a scheme is practical: how each scheme performs on CPU versus \ac{GPU}, how its throughput scales across CPU cores, and how expensive it is to build the index from scratch  (\Cref{sec:evaluation}).
    \item We release all results underlying every experiment, so the comparison is fully transparent and  reproducible.
\end{itemize}

\section{Background}
\label{sec:background}
We first outline how vector search operates, and then introduce our threat model.
We then discuss the four state-of-the-art schemes for privacy-preserving vector search we include in our benchmark, which span a wide range of privacy guarantees, and an unencrypted baseline.

\subsection{Vector search and \ac{IVF}}
\label{sec:bg-ivf}

\paragraph{Vector search}
A vector-search service stores a corpus of $N$ vectors with dimensionality $d$.
These vectors are typically embeddings that are generated from documents, \eg, using embedding models such as sentence transformers~\cite{reimers2019sentence} for textual data or image encoders for visual data~\cite{radford2021learning}.
Given a query vector $q$ issued by a client, the service returns the $k$ items most similar to $q$, a process known as \ac{NNS}~\cite{nene1997simple}.
Typical similarity distances used in vector search include L2 (Euclidean) distance, inner product (dot product), and cosine similarity.
Contemporary vector databases can contain billions of vectors, which makes an exact search, \eg, a brute-force query that scans all $N$ vectors with $O(Nd)$ compute cost per query, prohibitively slow.
Therefore, the majority of vector-search services rely on approximate \ac{NNS} where retrieval is approximate~\cite{indyk1998approximate}.
With approximate \ac{NNS}, an index built over all vectors narrows the search to a subset of candidates that, with high probability, contains the true nearest neighbors.

\paragraph{Vector search using IVF}
Modern unencrypted retrieval increasingly relies on graph indexes such as \textsc{HNSW}~\cite{malkov:2020:hnsw} and \textsc{DiskANN}~\cite{subramanya:2019:diskann}, which are often the faster choice on \ac{CPU}.
These indexes traverse the data along a query-dependent path, inspecting each candidate to decide which edge to follow next.
That data-dependent inspection is exactly what end-to-end encryption must hide.
As a result, the three schemes that encrypt the vectors end to end (\tiptoe, \emvp, and \bntm) cannot use graph indexes at all.
Since our unencrypted baseline and \sap could use them, but the other three schemes cannot, using graph indexes would complicate the comparison: any speed difference would mix the index choice with the cryptographic primitive used.

We therefore run all five schemes over a shared \ac{IVF} index~\cite{sivic:2003:videogoogle,jegou:2011:pq} instead, which is a commonly-used index family that all five schemes can natively support.
An \ac{IVF} index works as follows.
An offline $k$-means algorithm first partitions the vectors into roughly $\sqrt{N}$ clusters.
At query time, the cluster centroids are sorted by distance to $q$, the closest \texttt{nprobe} of them are picked, and the search analyzes only the distance to vectors inside those clusters.
The \texttt{nprobe} parameter trades off recall with computational cost, since larger values touch more clusters and recover more true neighbors at the cost of performing additional distance comparisons.
The \ac{IVF} per-cluster scan is embarrassingly parallel and therefore maps cleanly onto hardware accelerators such as \ac{GPU}s.

\begin{figure}[t]
  \centering
  \inputplot{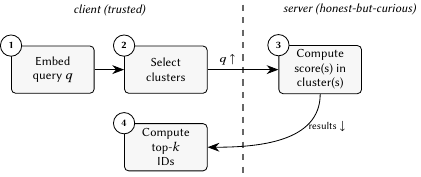}
  \caption{The four steps performed by the client and server in our per-query pipeline, shared by all five schemes, for privacy-preserving vector search using \ac{IVF}. %
    }
  \label{fig:schematic-pipeline}
\end{figure}

In this work, we assume a client holding a corpus.
This client first generates corpus embeddings and then uploads these embeddings to a server, which will perform the vector search on these corpus embeddings.
\Cref{fig:schematic-pipeline} sketches the pipeline when a client issues a query $q$ to the server, which consists of four steps: (1) the client \emph{embeds} its query $q$ using the same embedding model as used when creating the corpus embeddings, (2) a \emph{routing} step picks the \texttt{nprobe} target clusters, (3) the server \emph{scores} the candidates inside those clusters, and (4) the top-$k$ IDs are returned to the client.
The routing decision, \ie, computing distances to centroids and picking the \texttt{nprobe} closest clusters to analyze, is performed by the client, who holds the plaintext query and cluster centroids. %
We remark that all schemes, except \tiptoe, reveal to the server which clusters a query targets.
Thus, the server learns \texttt{nprobe} cluster indices for every query even though it never sees the query vector itself when using \emvp and \bntm.
The scoring step is where the schemes diverge most in what they leak to the server, ranging from the query, corpus, and scores in plaintext down to nothing beyond the routed cluster index.

\subsection{Threat model}
\label{sec:bg-threat}
By default, we assume an honest-but-curious server that faithfully executes the vector search mechanism but exploits its observations to infer as much private information as it can.
It is persistent: it retains every query it observes, along with the relative timing between queries, indefinitely.
The server stores the corpus vectors and associated \ac{IVF} indexes. %
The client, however, is fully trusted: it does not leak its own queries, and it safeguards any secret key material the scheme issues to it (\eg, the symmetric key used by \sap).
We note that \bntm also defends against a server that deviates from the protocol by returning erroneous responses; we further discuss this property in \Cref{sec:bg-schemes}.

The four cryptographic schemes differ in whether their privacy guarantee, \ie, bounding what the server learns to some specified leakage, holds up under repeated observation.
\tiptoe, \emvp, and \bntm hold up indefinitely: query after query, the server learns nothing beyond the system's specified leakage.
\sap does not: even a single view of the perturbed corpus reveals the relative distances between points.
Repeated queries make this worse: each query is perturbed independently, so a server observing many queries toward the same region of the embedding space can average away the noise and have precise insight into the access patterns of the users. However, noise-based schemes like \sap provide mitigation against inversion attacks \cite{tsai2026concept}, which makes them useful in production when only the raw contents of the documents are sensitive.
The geometry still reveals which documents lie close to each other.
A server that knows or guesses the embedding model can label these clusters with topics by matching them against embeddings of public text~\cite{naveed:2015:inference,grubbs:2017:leakage}.

\begin{table*}
	\centering
	\caption{The five schemes considered in this benchmark and their privacy guarantees. \emph{Security basis}: whether confidentiality reduces to keeping a key secret
		or to the hardness of a computational problem.
        \emph{Assumption}: the specific hardness assumption, where applicable.
        \emph{Leakage bounded under repeated queries}: whether the server learns nothing beyond its specified leakage even after repeated queries.
		\emph{Hides \acs{IVF} access pattern}: whether the server cannot tell which \ac{IVF} cluster a query targets.
        \emph{Malicious-server verifiability}: whether the client can verify that returned results match the requested computation.
        \emph{Graph-index compatible}: whether the scheme is compatible with graph-based \ac{ANN} indexes.}
	\label{tab:scheme-axes}
	\begin{tabular}{lccccc}
		\toprule
		& \plaintext{} & \sap{} & \emvp{} & \bntm{} & \tiptoe{} \\
		\midrule
		Security basis       & --       & key secrecy  & \multicolumn{3}{c}{computational hardness}  \\
        Assumption                                  & --         & --         & dual codes & LPN & LWE \\
		Leakage bounded under repeated queries        & \redcross  & \redcross  & \greencheck & \greencheck & \greencheck \\
		Hides \acs{IVF} access pattern   & \redcross  & \redcross  & \redcross  & \redcross  & \greencheck \\
		Malicious-server verifiability  & \redcross  & \redcross  & \redcross  & \greencheck & \redcross   \\
		Compatible with graph-based index  & \greencheck & \greencheck & \redcross  & \redcross  & \redcross   \\
		\bottomrule
	\end{tabular}
\end{table*}

\subsection{The five schemes under comparison}%
\label{sec:bg-schemes}

\begin{figure*}[t]
  \centering
  \inputplot{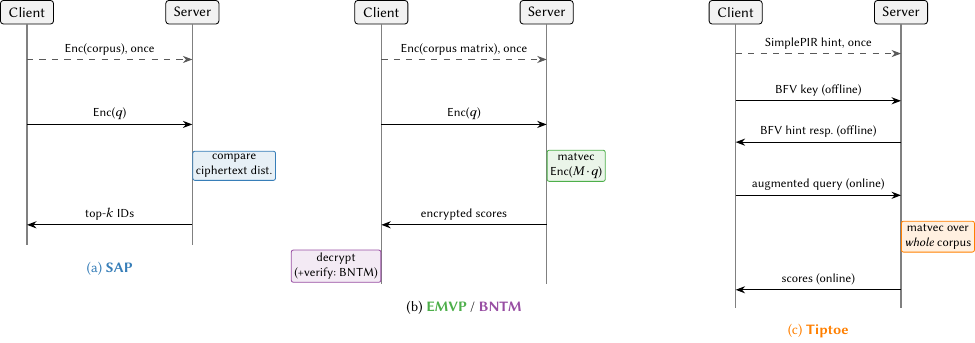}
  \caption{Communication between the client and server, and the performed operations during each of the four privacy-preserving vector search mechanisms, with time running from top to bottom. Dashed arrows indicate one-time setup operations and solid arrows indicate operations performed for each query. %
  }
  \label{fig:schematic-protocols}
\end{figure*}

We next discuss the five state-of-the-art schemes included in our comparison, which includes four privacy-preserving vector search mechanisms and a plaintext, unencrypted baseline.
The four privacy-preserving schemes rely on two different kinds of security argument.
\sap uses \emph{symmetric encryption}: the client holds a secret key and applies a keyed scaling to each vector, so security reduces to keeping that key secret from the untrusted server; unlike \emvp and \bntm, the resulting ciphertexts are not indistinguishable from random.
The other three privacy-preserving schemes (\emvp, \bntm and \tiptoe) instead rely on a \emph{computational hardness assumption}: each approach is secure only so long as a particular math problem stays infeasible to solve, which no one has proven.
Those problems belong to the \emph{learning} family, which all state that recovering a hidden linear function from noisy observations is computationally infeasible.
They differ in the algebraic setting and in how long the assumption has been studied: \emvp relies on \emph{secret dual codes}, \bntm on \acf{LPN}~\cite{blum:1994:lpn}, and \tiptoe on \acf{LWE}~\cite{regev:2009:lwe}.
We now outline each of these schemes, and also visualize the steps between the client and server within each privacy-preserving scheme in \Cref{fig:schematic-protocols}.
\Cref{tab:scheme-axes} also summarizes the five schemes and their privacy guarantees.

\paragraph{\plaintext{}.}
This is the baseline and default setting.
Vectors are stored and searched in the clear, without cryptographic protection.

\paragraph{\sap~\cite{fuchsbauer:2021:dcpe}.}
The main idea of \sap is to use a \ac{DCPE} scheme that lets the server compare distances between encrypted vectors directly, without ever decrypting them, also see \Cref{fig:schematic-protocols}(a).
The client holds a secret key and applies the same encryption to every vector it handles: it scales the vector by a per-key factor and adds a small, freshly sampled random perturbation.
It encrypts the corpus once, at index-build time, and uploads only the ciphertexts to the server.
At search time, the client encrypts each query the same way.
We note that scaling the vectors alone would be exactly distance-preserving but insecure, since an attacker could perfectly recover the corpus geometry.
The perturbed ciphertexts still approximately preserve the relative distances between points, up to controlled noise, the server can compare ciphertexts directly.
It routes \ac{IVF} queries and scans candidate clusters using the same code paths it would use on plaintext vectors, without ever decrypting anything.
A privacy parameter $\beta$ controls the magnitude of the perturbation added to each vector: a larger value of $\beta$ blurs the distance order more aggressively, yielding better privacy but lower recall.
Smaller values of $\beta$ approach the recall of the plaintext baseline (also see \Cref{sec:eval-privacy}).

\paragraph{\emvp~\cite{benhamouda:2025:emvp}.}
This scheme has been designed for settings with an honest-but-curious server threat model and hides both the query and the corpus contents from the server by computing entirely under encryption, also see \Cref{fig:schematic-protocols}(b).
The client encrypts the corpus once, as a matrix of ciphertexts, and uploads it to the server.
\ac{IVF} cluster routing still happens in the clear, so the query only needs to be encrypted with respect to the routed cluster's block.
At query time, the client sends an encrypted query vector, and the server answers by computing a \emph{matvec}, an encrypted matrix-vector product between the routed cluster's ciphertext matrix and the encrypted query, entirely under encryption; the result is returned to the client, who decrypts it to obtain the cluster's inner-product scores.
\emvp comes with the strong guarantee that to the server, the ciphertexts are indistinguishable from random, so it learns neither the query vector, the corpus vectors, nor the inner products themselves, and this guarantee holds indefinitely under repeated observation.
However, since cluster routing happens in the clear, the server does learn which cluster each query targets.

\paragraph{\bntm~\cite{braverman:2026:bntm}.}
This scheme is, from a high-level perspective, comparable to \emvp since it also encrypts vectors into a matrix, uploads them to the server, and then computes the distances under encryption.
There are two key differences.
Firstly, \bntm relies on the \acf{LPN} assumption, which is a long-studied assumption that gives a more conservative security argument than the newer one \emvp uses (which is based on dual codes)~\cite{blum:1994:lpn}.
Secondly, the client can verify, after receiving the encrypted response, that the server truthfully computed the requested inner products rather than fabricating them, with overwhelming probability and at a small per-query cost.
This check uses \emph{Freivalds' algorithm}, which is a randomized algorithm used to verify that the result of a matrix multiplication is correct~\cite{freivalds1979fast}.
We quantify this verification cost in \Cref{sec:eval-privacy}.

\paragraph{\tiptoe~\cite{henzinger:2023:tiptoe}.}
In contrast to the previous schemes discussed, \tiptoe also hides which cluster a query touches by having the server compute over the \emph{entire} corpus for every query, rather than just the routed cluster, also see \Cref{fig:schematic-protocols}(c).
Cluster routing itself stays client-side: the client scans the public centroids locally to pick its target cluster, and that choice never leaves the client.
It then builds an \emph{augmented query} that spans all clusters, carrying the query vector in the target cluster's block and zeros everywhere else, and encrypts the whole vector under \ac{LWE}~\cite{regev:2009:lwe}.
The server answers it with \textsc{SimplePIR}~\cite{henzinger:2023:simplepir}, a single-server \ac{PIR} scheme that replies to an encrypted query by multiplying it against the entire database, \ie, a single matvec of the \emph{whole} corpus matrix against the ciphertext.
Its per-query work is therefore proportional to the whole corpus rather than to the target cluster.
The client also needs a \textsc{SimplePIR} \emph{hint} to decode the result; because the hint depends only on the corpus and not the query, this exchange (handled under \ac{BFV} homomorphic encryption~\cite{fan:2012:bfv}) runs \emph{offline}, ahead of the query, leaving only the \emph{online} matvec on the critical path.
Because the augmented query is encrypted, the server cannot tell which block is non-zero, so it learns nothing about which cluster was targeted; the decrypted result carries the inner-product scores for clusters chosen by the client alone.
\tiptoe thus has the strongest privacy guarantees of all schemes by hiding both the vector contents and the cluster choice, but comes with very high per-query computational cost.

\section{Benchmark setup}
\label{sec:methodology}
We next describe the setup of our benchmark, and provide the implementation details of our five included schemes.

\paragraph{Workload.}
As corpus, we generate vectors from the \msmarco passage collection~\cite{nguyen:2016:msmarco}, which contains 8\,841\,823 passages.
This dataset is frequently used to evaluate vector-based retrieval in \ac{RAG} settings.
We embed each passage with the \efive embedding model~\cite{wang:2022:e5} into one 768-dimensional vector.
During our experiments, we consider two corpus sizes: a 100\,k-passage subset, referred to as \msmarcosmall, and the full 8.8\,M-passage collection, used when the scale of the corpus would affect the target metric, referred to as \msmarcofull.
We also deterministically sample 1\,000 questions from the \msmarco dataset and use these as query workload.
By default, we report recall@10, \ie, the fraction of a query's ten true nearest neighbors
that appear among the ten returned IDs: $|R_{10} \cap G_{10}| / 10$, where $R_{10}$ are the
returned IDs and $G_{10}$ the ground-truth neighbors.
To obtain $G_{10}$, we perform a brute-force search using L2 distance in the unencrypted corpus.
As similarity metric, we use L2 distance for \plaintext and \sap, and inner product for \emvp, \bntm, \tiptoe.
Since \efive embeddings are unit-normalised, the top-$k$ ranking induced by L2 distance is identical to the one induced by inner product, which lets us evaluate all schemes against a single ground truth despite their differing similarity metrics.

\paragraph{\ac{IVF} configuration.}
Every scheme implements the same scoring interface and reads the same \ac{IVF} partition for a given query.
We set the number of clusters $n_{\text{centroids}} $ to $\lceil \sqrt{N} \rceil$, yielding 317 clusters for \msmarcosmall and 2\,967 clusters for \msmarcofull.
To assign vectors to clusters, we run $k$-means with a fixed seed of $42$ for $25$ iterations of Lloyd's algorithm~\cite{lloyd:1982:kmeans}, which is the standard k-means loop that alternates nearest-centroid assignment with centroid recomputation.
The assignment of every vector to one of the $\lceil \sqrt{N} \rceil$ clusters, is computed once on the plaintext vectors and consistent across all five schemes.
This yields a mean cluster size $m = N/c \approx 315$ for \msmarcosmall and $m \approx 2\,966$ for \msmarcofull.
The schemes using cryptography encrypt the vectors only after they have been assigned to clusters, so encryption never changes the assignment.
This makes a cross-scheme comparison clean: because the \ac{IVF} index is identical for every scheme, any measured difference originates from the specifications of the underlying cryptographic scheme.

\paragraph{Per-scheme parameters.}
We configure each cryptographic scheme for \emph{128-bit security} (\ie, $\lambda = 128$).
For \emvp and \tiptoe, we pick parameters so that the best known attack on the hardness assumption the scheme rests on costs about $2^{128}$ operations, which is a standard cryptographic target and on the same level as AES-128.
For \bntm, whose paper does not provide concrete \ac{LPN} parameters, we instantiate $\lambda = 128$ through the parameter heuristic of the original paper; we could not certify the concrete security level of the resulting tuple.
Among the parameter sets that meet these requirements, we take the lowest-overhead one.
We provide all per-scheme parameters in \Cref{tab:scheme-params}.%

Two different $128$s appear in the paper and should not be confused.
The first one is $\lambda = 128$ as discussed above: it bounds \emph{privacy}, so breaking confidentiality costs about $2^{128}$ operations.
This applies to \emvp, \bntm, and \tiptoe.
The second one is a \emph{soundness} bound specific to \bntm: its iterated-Freivalds verifier (used to verify the server response, see \Cref{sec:bg-schemes}) runs $\lambda' = 3$ trials per query and therefore fails to catch a cheating server with probability at most $2^{-128}$.
\sap is the exception on both counts: it uses a 128-bit \emph{symmetric key}, but its privacy against an honest-but-curious server comes from the distance-distorting perturbation $\beta$, not from a computational-hardness assumption.
We evaluate different values of $\beta \in \{0, 0.25, 0.5, 0.75, 1\}$ and quantify the impact on recall in \Cref{sec:eval-privacy}.
Unless stated otherwise, we evaluate \sap with $\beta = 0$.

\paragraph{Implementation deviations.}
Our implementation differs from the original papers discussing the included schemes in two places.
First, \tiptoe normally refines its clustering in two ways: it recursively splits oversized clusters to keep them balanced, and it assigns the 20\,\% of vectors nearest a cluster boundary to two clusters instead of one.
We turn off both features so that every scheme uses the same \ac{IVF} partitions.
We validated our \tiptoe implementation against the official Go implementation~\cite{henzinger:2023:tiptoe} (patched to match this change), and found that the two implementations give identical results, on both recall and the per-query top-$k$ result IDs.

Second, \bntm has no public reference implementation.
Thus, we follow the paper's parameters~\cite{braverman:2026:bntm} and implementation details to the best of our ability, but stop short of one refinement.
The paper's mask is \emph{trapdoored}, meaning it carries a secret shortcut that lets the client undo it faster than a generic mask would allow.
We implement that mask, and we decode through the shortcut it provides: the client removes the mask with one small dense product plus one sparse product, rather than with the full $O(mn)$ multiplication a generic mask would force.
What we do not implement is the paper's \emph{recursive} refinement, which applies the same trick a second time inside the mask's own dense factor.
Our client unmasking therefore saves a constant factor over naive unmasking, where the full construction is sublinear in $mn$, so the \bntm client-side latency we report is still an upper bound.
This is a performance caveat, not a privacy break: hiding the corpus still reduces to \ac{LPN}, and the missing refinement leaves the server-side computation, the communication bytes, the recall, and the $2^{-128}$ malicious-server detection all unchanged.
We also cannot certify the exact security level, since the paper does not provide the exact \ac{LPN} parameters.

\paragraph{Hardware.}
We run all experiments on a single machine equipped with a dual-socket Intel Xeon Gold 6426Y, \num{32} physical cores plus \ac{SMT} for 64 logical cores and with \SI{128}{GB} of system memory.
This machine also contains an NVIDIA RTX 5000 Ada GPU (with \SI{32}{GB} of video RAM) which we leverage for our experiments that use a \ac{GPU}.
The only thing that changes between a \ac{CPU} and a \ac{GPU} run is where the server-side scoring code runs (\ie, computing the distances between the query and vectors in each cluster, step 3 in \Cref{fig:schematic-pipeline}).

\paragraph{Measurement.}
We apply a few standard controls so the timings are stable and reproducible.
All experiments use the 64 logical cores by default; \Cref{sec:eval-deploy} shows that 32 threads would be slightly faster, a choice that affects all schemes alike.
\bntm's response verification is disabled by default, since a malicious server is outside our default threat model; we enable it only in the verification experiment of \Cref{sec:eval-privacy}.
We set every machine's \ac{CPU} frequency governor to \texttt{performance}, which holds the cores at their top clock instead of letting the frequency ramp up and down, so latency does not change with the processor's power state.
We also drop the operating system's page cache between runs so each run starts cold and no scheme benefits from data another left in memory.
When changing the number of threads in \Cref{sec:eval-deploy}, we pin threads to socket 0 up to 16 threads and run larger counts unpinned: on our dual-socket machine the memory is split between the two sockets (a \ac{NUMA} layout) and reaching the other socket's memory is slower, so confining threads to one socket lets us attribute the efficiency drop in our parallel throughput experiments to real per-core contention rather than to stray cross-socket traffic.
Finally, every figure uses the \emph{realised} per-query communication cost, \ie, the actual bytes that left or arrived for that query, not the per-cluster mean over the whole index; this distinction matters for clustered corpora whose cluster sizes vary by an order of magnitude (we quantify this in \Cref{sec:eval-comm}).
Unless stated otherwise, every experiment uses all 64 logical CPU cores within each query, and queries are issued sequentially; only the parallel-scaling sweep (\Cref{sec:eval-deploy}) varies the thread count.

\paragraph{Reproducibility.}
All artifacts are made public\footnote{See \url{https://github.com/sacs-epfl/secure-vector-search}.} and every figure in this paper can be reconstructed from the raw measurements.
We repeat every experiment three times and report mean values.
 
\section{Benchmark results}
\label{sec:evaluation}
Our benchmark answers the following main question: \emph{at the operating points that matter to a vector-search deployment, what does cryptographic privacy cost in terms of computational and communication cost, and search accuracy?}
We answer this through the following sub-questions:

\begin{itemize}
    \item What is the trade-off between throughput and recall for all schemes included in the benchmark, on different corpus sizes and on a CPU and GPU (\Cref{sec:eval-pareto})?
    \item How much does the GPU speed up the different schemes in terms of throughput and query latency (\Cref{sec:eval-cpu-gpu})?
    \item What is the (per-query) online, offline and (one-time) setup communication cost incurred by each scheme (\Cref{sec:eval-comm})?
    \item What is the end-to-end query latency of each scheme and how much do the different operations contribute to this latency (\Cref{sec:eval-latency})?
    \item How does the perturbation strength $\beta$ in \sap affect recall, and what is the compute overhead of response verification by the client in \bntm (\Cref{sec:eval-privacy})?
    \item How much does multi-threading affect per-query latency of each scheme (\Cref{sec:eval-deploy})?
    \item How long does it take for each scheme to build the index by the client (\Cref{sec:eval-index-build})?
\end{itemize}

\subsection{The throughput-recall trade-off}
\label{sec:eval-pareto}
\begin{figure*}
	\centering
	\inputplot{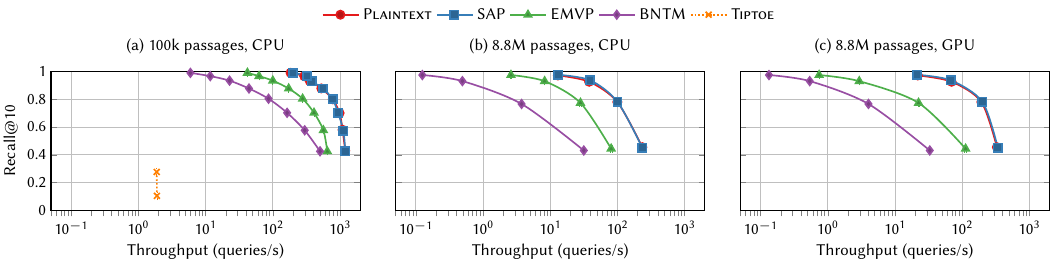}
	\caption{Recall@10 vs.\ sustained throughput (with a logarithmic horizontal axis). \textbf{(a)} \msmarcosmall on \ac{CPU}; \textbf{(b)} \msmarcofull on \ac{CPU}; \textbf{(c)} \msmarcofull on \ac{GPU}. \tiptoe{} is out of frame in (b) and (c), and runs at its own recall regime ($\approx$ 0.11--0.28), so its curve sits low by construction.
    }
	\Description{Three side-by-side line plots of recall versus sustained throughput on a log-scale x-axis. Chart (a) shows five curves at the 100k validation corpus — Plaintext, three cryptographic IVF variants, and Tiptoe at low recall. Charts (b) and (c) show four curves each (all schemes except Tiptoe) at the 8.8M corpus on CPU and GPU respectively.}
	\label{fig:recall-throughput}
\end{figure*}
Our first experiment measures the trade-off between throughput (the number of answered queries/s) and recall@10 for all five schemes and for \msmarcosmall and \msmarcofull.
For all schemes, except \tiptoe, we vary the \texttt{nprobe} parameter (\ie, the number of clusters to check), which trades compute cost for recall.
Since \tiptoe is architecturally different, we evaluate \tiptoe with two quantization levels: 3 and 4 bit.
Since we are interested in the per-query cost, we measure throughput by issuing queries sequentially and then take the reciprocal of mean single-query latency.
Here, we focus on the throughput-recall trade-off and later quantify the communication cost and end-to-end latency in \Cref{sec:eval-comm} and \Cref{sec:eval-latency}, respectively.

\Cref{fig:recall-throughput}(a) shows the throughput-recall trade-off on \ac{CPU} for \msmarcosmall (all five schemes), and \Cref{fig:recall-throughput}(b) for \msmarcofull (all schemes except \tiptoe), both with a logarithmic horizontal axis.
\Cref{fig:recall-throughput}(c) repeats the \msmarcofull measurements on the \ac{GPU} and is discussed in \Cref{sec:eval-cpu-gpu}.
\Cref{fig:recall-throughput}(a) considers eight values of \texttt{nprobe}, powers of two from 1 to 128.
The \msmarcofull runs in (b) and (c) sweep \texttt{nprobe} over 1, 8, 64 and 256.
In both settings, increasing the \texttt{nprobe} parameter increases recall at the cost of throughput.
\Cref{fig:recall-throughput}(a) shows that for the \bntm scheme, for example, recall@10 increases from 0.43 for $nprobe=1$ to 0.99 for $nprobe=128$.
At the same time, throughput for \bntm drops from 506 queries/s to 6.0 queries/s when increasing $nprobe$ from 1 to 128, an 84$\times$ decrease.
We observe similar trends for the other schemes, highlighting the recall-throughput trade-off.

We also observe in \Cref{fig:recall-throughput} that the schemes separate into distinct tiers whose gaps widen as the privacy guarantee strengthens.
The recall and throughput of \sap is very close to that of \plaintext across every operating point on every chart.
In other words, the privacy offered by \sap essentially adds no compute overhead because the server side runs the same scoring computation as \plaintext but on shuffled ciphertexts.
The throughput performance of \emvp at a recall of $\approx$0.9 is 3.5$\times$ below that of \plaintext with \msmarcosmall and 4$\times$ below with \msmarcofull.
The throughput of \bntm sits below \emvp at a recall of $\approx$0.9: 4.4$\times$ lower throughput (100k, recall@0.9) with \msmarcosmall and 44.6$\times$ lower throughput (8.8M, recall@0.9) with \msmarcofull.
Growing the corpus from 100\,k to 8.8\,M passages preserves the ordering of the schemes (\Cref{fig:recall-throughput}(a) vs.\ \Cref{fig:recall-throughput}(b)).

The throughput of \tiptoe on the small corpus (1.9 queries/s) is far below all other schemes, two orders of magnitude below \plaintext, and also comes at a much lower recall.
\tiptoe is not plotted in \Cref{fig:recall-throughput}(b) and (c) because it is too slow at that scale: its per-query cost grows linearly in corpus size, which extrapolates the measured \msmarcosmall latency to \SI{46}{s} per query on \msmarcofull, resulting in a throughput of 0.02 queries/s.
\tiptoe's low recall is structural for both evaluated quantization levels and has two separate explanations.
First, \tiptoe searches exactly one \ac{IVF} cluster per query (the server never learns which), so any ground-truth passage that lies outside that cluster can never be returned; on our corpus only 44\,\% of a query's ground-truth top-10 passages fall in the cluster the query is routed to.
Second, \tiptoe's encrypted matvec operates on low-precision integers, so the embeddings must be quantised before scoring, which reorders passages within the one cluster that is searched.
Together they leave recall@10 at 0.11 at 3-bit precision and 0.28 at 4-bit precision.
These figures match the authors' reference implementation within 0.1 percentage point ($0.110$ vs $0.111$ at 3-bit), and we hold the clustering layer identical across all five schemes rather than apply \tiptoe{}'s published recall-boosting extensions (recursive cluster splitting, boundary double-assignment), which would break the controlled comparison. \tiptoe{} therefore trades recall for sublinear communication, its declared design point; the \Cref{fig:recall-throughput} position quantifies that trade at benchmark scale.

\begin{takeaway}
Cryptographic indistinguishability costs a small factor, not orders of magnitude: \emvp runs 4$\times$ below \plaintext at recall 0.9. \sap comes with almost no computational overhead but reveals the approximate corpus geometry; the throughput gap between \sap and \emvp is the price of hiding that geometry.
\end{takeaway}

\subsection{Speedup by the GPU}
\label{sec:eval-cpu-gpu}
Next, we analyze how offloading the score computation to a GPU on the server affects the performance of all schemes, except \tiptoe.
We consider a GPU-native implementation of \tiptoe beyond the scope of our benchmark.

\paragraph{Recall-throughput}
\Cref{fig:recall-throughput}(c) shows the recall-throughput trade-off when using a \ac{GPU}.
Moving from \ac{CPU} to \ac{GPU} (\Cref{fig:recall-throughput}(b) vs. \Cref{fig:recall-throughput}(c)) helps them unevenly: at recall 0.9, \plaintext and \sap gain throughput (48.7 to 91.2 and 53.4 to 98.5 queries/s, respectively), \emvp loses throughput (12.0 to 6.6 queries/s), and \bntm barely moves (also see \Cref{tab:cpu-vs-gpu}).

This is because most of their per-query time is client-side work that the \ac{GPU} is unable to speed up.
\Cref{sec:eval-latency} discusses this further by presenting a latency breakdown of server and client operations.

\begin{figure}[t]
  \centering
  \inputplot{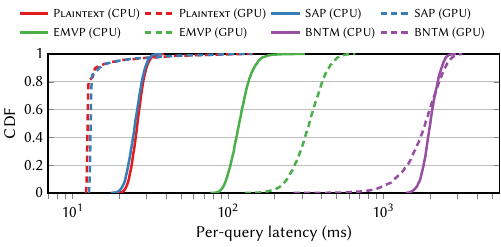}
  \caption{Per-query latency \ac{CDF} for all schemes, on CPU and GPU, on \msmarcofull at \texttt{nprobe} = 64 (recall $\approx$ 0.93). This is a slightly higher operating point than the recall of exactly 0.9 in \Cref{tab:cpu-vs-gpu}, so the throughput there is a bit higher than these latencies suggest.
  }
  \Description{Eight cumulative-distribution curves of per-query latency on a logarithmic x-axis, one per (scheme, device) pair. Plaintext and SAP GPU curves are roughly half a decade left of their CPU counterparts; EMVP GPU is right of CPU; the two BNTM curves overlay around the 2-second mark.}
  \label{fig:cpu-vs-gpu-latency-cdf}
\end{figure}

\paragraph{Effect on throughput}
\Cref{tab:cpu-vs-gpu} compares \ac{CPU} and \ac{GPU} performance at the same recall of 0.9.
The setup is comparable to the one used in \Cref{sec:eval-pareto}.
Since no \texttt{nprobe} setting lands on recall 0.9 exactly, the throughput values in \Cref{tab:cpu-vs-gpu} are obtained by taking, for each scheme, the two measured settings whose recalls lie just below and just above 0.9, and linearly interpolating the throughput between them.
\plaintext and \sap run 1.8--1.9$\times$ faster on the \ac{GPU} than on the 64-thread \ac{CPU} baseline.
\emvp loses throughput on the \ac{GPU} at this corpus size because its encrypted matrix is memory-bandwidth-bound, not compute-bound: it performs a single modular multiply per \SI{8}{B} ciphertext field element it reads, so the \ac{GPU} spends its time waiting on memory rather than computing.
Ciphertext expansion makes this worse: the encrypted matrix is 91\,GB for \msmarcofull, which doesn't fit in the 32\,GB of GPU memory.
Thus, the encrypted matrix cannot stay resident and streams cluster-by-cluster over the PCIe bus, which is much slower than the on-device memory the \plaintext and \sap score computation logic read from.
\bntm sits at single-digit throughput at recall 0.9 (1.09 qps on CPU and 1.18 qps on \ac{GPU}): its per-query time is dominated by client-side decoding (see \Cref{sec:eval-latency}), which the \ac{GPU} does not speed up.
\Cref{fig:cpu-vs-gpu-latency-cdf} complements \Cref{tab:cpu-vs-gpu} and shows the distribution of per-query latency at recall $\approx$ 0.9. %

\begin{table}[t]
  \centering
  \caption{Throughput (in queries/s, or qps) at recall@10=0.9, on \msmarcofull.
  \tiptoe is omitted since it never reaches that recall.}
  \label{tab:cpu-vs-gpu}
  \begin{tabular}{l S[table-format=2.2] S[table-format=2.2] S[table-format=1.2]}
    \toprule
    Scheme & {\ac{CPU} (qps)} & {\ac{GPU} (qps)} & {Speedup} \\
    \midrule
    \plaintext     & 48.65 & 91.17 & 1.87$\times$ \\
    \sap           & 53.40 & 98.50 & 1.84$\times$ \\
    \emvp          & 12.01 &  6.63 & 0.55$\times$ \\
    \bntm          &  1.09 &  1.18 & 1.08$\times$ \\
    \bottomrule
  \end{tabular}
\end{table}

The \plaintext and \sap \ac{GPU} curves sit left of their \ac{CPU} counterparts along the whole distribution, \emvp's sits right along the whole distribution, and \bntm's two curves overlay around \SI{2}{s}.
The \ac{GPU} runs do spread more (p95 at 1.5$\times$ the median, vs 1.2$\times$ on the \ac{CPU}), but even at p95 \plaintext and \sap stay 1.6$\times$ faster on the \ac{GPU}.
We attribute the wider spread to per-query kernel-launch and host-device synchronization jitter, which is large relative to the short \ac{GPU} kernels.

\begin{takeaway}
Offloading the score computation to the \ac{GPU} on the server pays off only where this computation is not memory-bound: \plaintext and \sap gain 1.8--1.9$\times$ in throughput, but \emvp and \bntm show little increase, or even performance degradation.
\end{takeaway}

\begin{table*}[t]
    \small
  \centering
  \caption{Per-query online and offline communication costs, and bytes read from memory by the server, on \msmarcosmall{} (top, \texttt{nprobe} = 32) and \msmarcofull{} (bottom, \texttt{nprobe} = 64). All schemes, except \tiptoe, obtain a recall@10\,$\approx 0.9$; the communication cost of \tiptoe is independent of its recall. All rows are end-to-end measurements except \tiptoe$^\dagger$ at 8.8\,M, whose costs are estimated from closed-form cost expressions (see \Cref{sec:appendix-comm-formulas}) since the \ac{LWE} matvec computation cannot run at this scale.
  }
  \label{tab:byte-budget-100k}\label{tab:byte-budget-8m}
  \begin{tabular}{l r | r r | r | r}
    \toprule
    & & \multicolumn{2}{c|}{Comm. / query (online)} & \multicolumn{1}{c|}{Offline / query} & \\
    \cmidrule(lr){3-4}
    Scheme & setup (one-time) & query $\uparrow$ & response $\downarrow$ & $\uparrow$ / $\downarrow$ & mem-read bytes / query \\
    \midrule
    \multicolumn{6}{l}{\emph{\msmarcosmall{} (\texttt{nprobe}\,=\,32)}} \\
    \plaintext  & ---           & \SI{3}{KB}   & \SI{80}{B}    & ---                          & \SI{31}{MB}   \\
    \sap        & \SI{614}{MB}  & \SI{6}{KB}   & \SI{80}{B}    & ---                          & \SI{31}{MB}   \\
    \emvp       & \SI{1.0}{GB}  & \SI{10}{KB}  & \SI{7.2}{MB}  & ---                          & \SI{104}{MB}  \\
    \bntm       & \SI{819}{MB}  & \SI{8}{KB}   & \SI{94}{KB}   & ---                          & \SI{83}{MB}   \\
    \tiptoe{}   & \SI{16}{MB}   & \SI{1.9}{MB} & \SI{7.8}{KB}  & \SI{42}{MB} / \SI{164}{KB}   & \SI{16}{MB}   \\
    \midrule
    \multicolumn{6}{l}{\emph{\msmarcofull{} (\texttt{nprobe}\,=\,64)}} \\
    \plaintext         & ---             & \SI{3}{KB}   & \SI{80}{B}    & ---                          & \SI{636}{MB}  \\
    \sap               & \SI{54}{GB}     & \SI{6}{KB}   & \SI{80}{B}    & ---                          & \SI{636}{MB}  \\
    \emvp              & \SI{91}{GB}     & \SI{10}{KB}  & \SI{126}{MB}  & ---                          & \SI{2.15}{GB} \\
    \bntm              & \SI{72}{GB}     & \SI{8}{KB}   & \SI{1.6}{MB}  & ---                          & \SI{1.69}{GB} \\
    \tiptoe{}$^\dagger$ & \SI{151}{MB}    & \SI{18}{MB}  & \SI{74}{KB}   & \SI{42}{MB} / \SI{492}{KB}   & \SI{151}{MB}  \\
    \bottomrule
  \end{tabular}
\end{table*}

\subsection{Communication cost}
\label{sec:eval-comm}
Next, we quantify the communication cost of each scheme in our benchmark.
We particularly focus on the communication volume between the client and server, and distinguish between four types of traffic.
First, we measure the \emph{one-time setup cost} that encrypts and uploads the corpus at index time to the server, which is a cost that is amortised across every subsequent query.
Second, we measure \emph{online per-query traffic} (up- and download volume), \eg, the client uploading the query embedding and the server replying with encrypted scores.
Third, for \tiptoe we consider a per-query \emph{offline} round-trip that precedes the online \textsc{matvec} computation, in which the client during the offline phase sends a \ac{BFV}-encrypted secret key to the server and receives the \ac{BFV}-encrypted hint response from the server.
Fourth, we measure the volume of data that the server is reading from the memory per scoring call, which predicts the memory bandwidth requirement by the server.
We run the communication cost experiments on both \msmarcosmall and \msmarcofull, while ensuring that all schemes, except \tiptoe, achieve comparable recall@10\,$\approx 0.9$ by fixing \texttt{nprobe} = 32 and \texttt{nprobe} = 64, respectively.
\tiptoe runs at its own recall@10\,$\approx 0.28$ point (similar to \Cref{sec:eval-pareto}) but its recall does not influence the communication cost in our setup because this scheme always queries a single cluster.

\Cref{tab:byte-budget-100k} shows the four types of communication costs for each of the five schemes, for both \msmarcosmall and \msmarcofull.
The schemes fall into three communication shapes: \plaintext pays only the online round-trip; \sap, \emvp and \bntm add the one-time setup upload; \tiptoe additionally pays a per-query offline round-trip before the matvec computation.

\paragraph{Setup cost}
The setup column is where the privacy-preserving \ac{IVF} schemes' upfront tax lands.
\sap uploads the \ac{DCPE}-encrypted corpus, with a size of \SI{614}{MB} for \msmarcosmall; this is the cheapest compared to \emvp and \bntm because its ciphertexts keep the plaintext dimension $d$.
\emvp and \bntm both upload an $m \times n$ encrypted matrix per cluster over 317 clusters ($n > d$), resulting in a setup size of \SI{1.0}{GB} and \SI{819}{MB}, respectively.
These costs are one-time and amortise across subsequent queries.
\plaintext has no setup cost because no privacy constraint forces the corpus through the client: the server can ingest it directly.
If the corpus instead starts at the client, \plaintext would upload the raw vectors, \SI{307}{MB} for \msmarcosmall: half of \sap's setup cost, because \sap{} ciphertexts use \SI{8}{B} per coordinate instead of \SI{4}{B}.
\tiptoe's setup cost comprises the \SI{16}{MB} SimplePIR hint and is almost two orders of magnitude smaller than the setup cost for \emvp because the hint is $m_{\max} \times n_{\textrm{LWE}} \times \SI{8}{B}$ and $n_{\textrm{LWE}} = 2048$ is much smaller than $m \times n$ for the \ac{IVF} encrypted matrix used by \emvp and \bntm.
On \msmarcofull, the setup uploads of \sap, \emvp and \bntm grow linearly with $N$ (88$\times$) to \SI{54}{GB}, \SI{91}{GB} and \SI{72}{GB}, while \tiptoe's hint grows with the cluster size (9.4$\times$) to \SI{151}{MB}.

\paragraph{Per-query communication cost}
On \msmarcosmall, the per-query up- and download traffic varies significantly across schemes, between \SI{3}{KB} and \SI{1.9}{MB}, and \SI{80}{B} and \SI{7.2}{MB}, respectively.
\plaintext uploads a \SI{3}{KB} query to the server (a vector with 768 coordinates at \SI{4}{B} each) and \sap a \SI{6}{KB} query (the same 768 coordinates at \SI{8}{B} each); both return only the top-$k$ identifiers and scores, which accounts for \SI{80}{B} at $k = 10$.
\emvp and \bntm instead return one entry per corpus vector in every probed cluster, so their response grows with \texttt{nprobe} and with the cluster size, not with $k$.
\bntm returns a single field element, \SI{8}{B}, per corpus vector, with no secret-sharing layer: a probed cluster averages 368 vectors (\ac{IVF} routing favours clusters denser than the corpus mean of $m \approx 315$), which requires communicating \SI{2.9}{KB} to the client, and summing over the \texttt{nprobe} = 32 probed clusters gives \SI{94}{KB} per query.
\emvp returns a whole $s \times m$ block-product matrix per probed cluster (where $s = 76$ coded scalars per vector), which results in \SI{219}{KB} download traffic per cluster and \SI{7.2}{MB} once summed over the same 32 clusters.
This is a $76\times$ penalty over \bntm at comparable recall, because the client must recover each score from $s$ coded scalars instead of reading it directly.%
\tiptoe shows the opposite pattern: uploading the encrypted \ac{LWE} query to the server incurs \SI{1.9}{MB} (plus a \SI{42}{MB} \ac{BFV} offline-phase round-trip), while the \ac{LWE} response size is only \SI{7.8}{KB}.
Comparing the top and lower part of \Cref{tab:byte-budget-100k} shows how these costs scale with corpus size.
The query upload is unchanged for the four \ac{IVF} schemes, since it depends only on the vector dimension, but grows 9.5$\times$ for \tiptoe (\SI{1.9}{MB} to \SI{18}{MB}), tracking $\sqrt{N}$.
The responses of \emvp and \bntm grow 17$\times$, following the mean cluster size (9.4$\times$) and the doubled \texttt{nprobe}, while \plaintext and \sap still return only the \SI{80}{B} top-$k$.

\paragraph{Memory traffic}
Next, we analyze the per-query memory traffic required by the server to compute the scores, which is shown in the right-most column in \Cref{tab:byte-budget-100k}.
We measure this since the scoring computation on the server has to potentially read many clusters from memory which can become a bottleneck as the corpus size increases.
The score computation reads $O(m \times d_{\textrm{enc}})$ bytes from its own memory per cluster, where $d_{\textrm{enc}}$ is the scheme's effective row width (4-byte fp32 for \plaintext and \sap, $n_{\textrm{EMVP}} = 1292$ 8-byte u64 for \emvp{}, $n_{\textrm{BNTM}} = 1024$ 8-byte u64 for \bntm).
The encrypted schemes pay a 2.7--3.4$\times$ memory-traffic premium for the same recall \emph{even when} their wire-response is small (\eg, for \bntm), because direct delivery shrinks the \emph{returned scalars} but not the matrix the score computation must scan.
\tiptoe's matrix is the SimplePIR hint itself, which is \SI{16}{MB} for \msmarcosmall and \SI{151}{MB} for \msmarcofull.

The differences in memory-bandwidth and network requirements matter when selecting a scheme for deployment: network cost determines what deployment costs on a constrained or metered link, whereas memory bandwidth is what bounds server throughput and what determines whether moving the scoring computation to a \ac{GPU} pays off.
Choosing a scheme based on a low network volume footprint alone therefore mispredicts throughput.
\bntm is an example of this: it has the smallest response size of all cryptographic schemes (\SI{94}{KB}) but its in-memory scan cost is comparable to that of \emvp (\SI{83}{MB} vs \SI{104}{MB}).
The same gap pattern holds for \msmarcofull (\Cref{tab:byte-budget-8m}): \bntm's in-memory cost rises to \SI{1.69}{GB} vs \emvp's \SI{2.15}{GB}.

\Cref{sec:appendix-comm-formulas} derives closed-form expressions that yield the wire and in-memory byte counts for all schemes.

\begin{takeaway}
Based on network cost and memory bandwidth, the schemes rank differently: \bntm is the cheapest cryptographic scheme in terms of network volume yet reads as much memory as \emvp. %
\end{takeaway}

\begin{figure*}[t]
  \centering
  \inputplot{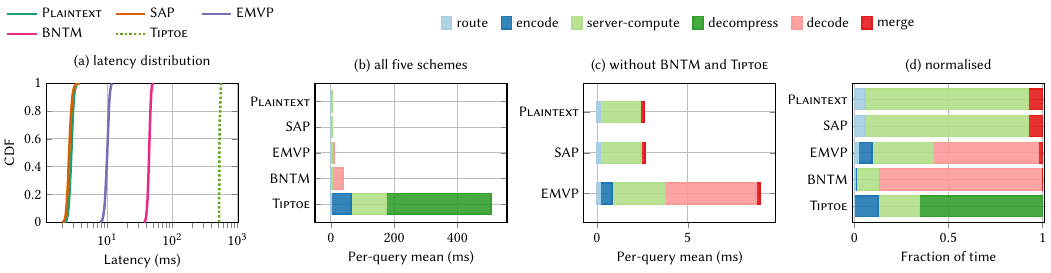}
  \caption{Per-query latency on \msmarcosmall.
  \textbf{(a)} The distribution of E2E query latency for each of the five schemes; all schemes except \tiptoe{} are at the recall $\approx 0.9$ operating point, \tiptoe{} at its own recall $\approx 0.28$ point (\Cref{sec:eval-pareto}).
  \textbf{(b)} The mean latency breakdown into the six operations for all five schemes;
  \textbf{(c)} the same without \tiptoe and \bntm, for readability;
  \textbf{(d)} the breakdown of (b) normalised to the total per-query time of each scheme.
  }
  \Description{Four panels. (a) Five CDFs of per-query latency on a logarithmic x-axis; the Plaintext, SAP, EMVP, and BNTM curves rise in the millisecond range, the Tiptoe curve far to the right. (b) Stacked horizontal bars of total per-query latency split by operation for all five schemes; the Tiptoe bar dominates, driven by decompress. (c) The three fastest schemes at a finer scale, where EMVP is decode-dominated and Plaintext and SAP are server-compute-dominated. (d) The same bars normalised to equal width, showing each scheme's composition.}
  \label{fig:latency-cdf}\label{fig:substeps-absolute}\label{fig:substeps-normalised}
\end{figure*}

\subsection{E2E latency and breakdown}
\label{sec:eval-latency}

\paragraph{Distribution of E2E latencies}
Next, we quantify the per-query E2E latency for all schemes, whose distribution is shown in \Cref{fig:latency-cdf}(a).
In line with previous experiments, all schemes, except \tiptoe, are evaluated at recall@10 $\approx 0.9$ but \tiptoe operates at a lower recall\,$\approx 0.28$ point.
We measure the per-query latency of \bntm with server response verification disabled.
\Cref{fig:latency-cdf}(a) shows that \plaintext, \sap and \emvp have the lowest end-to-end latency: a median latency of \SI{2.8}{ms} for \plaintext and \SI{2.6}{ms} for \sap, and \SI{10.0}{ms} for \emvp, with p99 below \SI{12}{ms}.
\bntm and \tiptoe have higher end-to-end latencies, with a median latency of \SI{44}{ms} and \SI{528}{ms}, respectively.
We observe that \tiptoe's \ac{CDF} curve compared to those of the other scheme shows the lowest variance on the horizontal axis because its per-query budget is dominated by SimplePIR's deterministic matvec computation, which runs in consistent time.

\paragraph{Latency breakdown}
We further analyze query latency by recording for each scheme the time spent in each of the following six operations: \emph{encode} (the client transforms or encrypts the query, step 1 in \Cref{fig:schematic-pipeline}), \emph{route} (the client scores the query against the centroids and picks the \texttt{nprobe} clusters to probe, step 2 in \Cref{fig:schematic-pipeline}), \emph{server-compute} (the server scores the probed clusters, step 3 in \Cref{fig:schematic-pipeline}), \emph{decompress} (the client unpacks the server response), \emph{decode} (the client recovers the scores from the response), and \emph{merge} (the client combines the per-cluster candidates into the global top-$k$, step 4 in \Cref{fig:schematic-pipeline}).
\Cref{fig:substeps-absolute}(b) shows the mean duration of each operation for each scheme.
For reference, \Cref{fig:substeps-normalised}(d) shows the same results as in \Cref{fig:substeps-absolute}(b) but normalizes each bar to 100\,\%. %
\tiptoe shows the highest total per-query mean latency of \SI{506}{ms}.
This latency is dominated by the \emph{decompress} operation (\SI{331}{ms} or 65\% of the total latency), followed by the latency required for server-compute (\SI{111}{ms}, 22\%) and encode (\SI{64}{ms}, 13\%).
\tiptoe's per-query latency goes mostly to decompressing the SimplePIR response.
\bntm has a total per-query mean of \SI{43}{ms}, which is dominated by the client-side \emph{decode} operation (\SI{35}{ms}, 81\,\%). %
We run \bntm without verification here  and analyze its verification cost in \Cref{sec:eval-privacy}.

The total per-query mean latency of the \plaintext, \sap and \emvp schemes are all under \SI{9}{ms}, and for presentation clarity we show these latency breakdowns separately in \Cref{fig:substeps-absolute}(c).
\plaintext and \sap both have a total per-query mean latency of \SI{2.6}{ms} and their latencies are dominated by the server-compute operation (86\,\%).
\emvp has a total per-query mean latency of \SI{9.0}{ms} and this latency is mainly attributed to the decode operation (56\,\%).

From \Cref{fig:substeps-absolute,fig:substeps-normalised} we conclude the following.
\plaintext and \sap are \emph{server-compute} bound: 86\,\% of their (small) compute time is spent in the scoring computation.
The three end-to-end-encrypted schemes are instead bound by \emph{client-side response processing}: \emvp and \bntm by \emph{decode} (dual-code decoding and trapdoor unmasking, respectively) and \tiptoe by decompressing the SimplePIR
response.
\begin{takeaway}
Each scheme's per-query cost is concentrated in a single pipeline segment: server-side scoring for \plaintext{} and \sap{}, client-side decoding for \emvp{} and \bntm{}, and response decompression for \tiptoe. Optimisation has one clear target per scheme.
\end{takeaway}

\subsection{Varying the privacy knobs}
\label{sec:eval-privacy}
\begin{figure}[t]
  \centering
  \inputplot{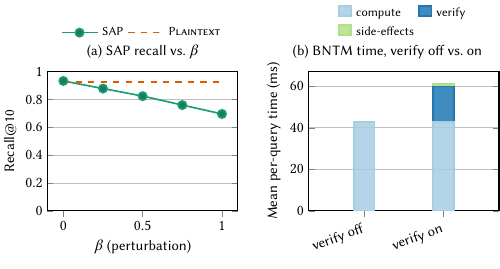}
  \caption{Privacy knobs in \sap and \bntm. (a) The effect of varying \sap{}'s perturbation strength $\beta$ from 0 (no privacy) to 1 on recall@10, on \msmarcofull. For reference, the recall@10 of \plaintext is 0.925 (dotted line). (b) \bntm per-query time breakdown at \texttt{nprobe}\,=\,32, with and without verification enabled, on \msmarcosmall.}
  \Description{Two charts side by side. (a) A line plot of SAP recall at beta 0, 0.25, 0.5, 0.75 and 1, descending as beta grows, with a horizontal line marking the plaintext baseline. (b) A stacked bar chart of BNTM per-query time with verification off vs on at nprobe=32: the off-bar shows compute only; the on-bar adds verify and side-effects segments above the same compute baseline.}
  \label{fig:privacy-knobs}
\end{figure}

\sap and \bntm expose configurable privacy parameters, and the two knobs trade against different axes.
\sap's perturbation strength $\beta$ yields stronger snapshot privacy at the cost of recall, and \bntm's verification toggle enables malicious-server detection at the cost of latency.
We report the effect of both these parameters on recall (for \sap) and on latency (for \bntm) in \Cref{fig:privacy-knobs}.

\paragraph{Perturbation $\beta$ in \sap}
\Cref{fig:privacy-knobs}(a) shows recall@10 for \sap when sweeping the perturbation strength $\beta$ from 0 (no perturbation) to 1, on the \msmarcofull dataset at \texttt{nprobe} = 64.
Recall@10 decreases roughly linearly with $\beta$: from 0.93 at $\beta = 0$ to 0.82 at $\beta = 0.5$ (under which close vectors become indistinguishable) and 0.70 at $\beta = 1$; this drop is the price paid for symmetric snapshot privacy.
At $\beta = 0$, recall closely matches the \plaintext baseline at the same \texttt{nprobe} (the horizontal dashed line which sits at recall@10 $= 0.925$), as expected: the scale-only transform in \sap preserves the ranking.
Note that \sap's throughput-recall curve in the \Cref{fig:recall-throughput} is computed with $\beta = 0$;
the recall cost of perturbing vectors with $\beta > 0$ is the drop shown in \Cref{fig:privacy-knobs}(a).
$\beta$ has no per-query latency cost: the perturbation is applied to the corpus at setup, so at a fixed \texttt{nprobe} the query-time work is identical for any $\beta$.

\paragraph{Toggling verification in \bntm}
\bntm includes a verification mechanism that protects against a malicious response by the server, \ie, the server cannot return a wrong answer on purpose without the client noticing, with detection probability $1 - 2^{-128}$ under the iterated-Freivalds verifier at $\lambda' = 3$ trials per query.
\Cref{fig:privacy-knobs}(b) shows the per-query time for \bntm on \msmarcosmall with \texttt{nprobe}=32.

We split this time into three parts: \emph{compute} (the full verify-off pipeline: the server computing scores in the cluster and client-side decoding), 
\emph{verify} (running the Freivalds verifier), and \emph{side-effects}, which captures the overhead the verify-on code path imposes on the rest of the pipeline.
The verify segment takes \SI{17}{ms} per query (40\,\% of the unverified \SI{43}{ms} baseline); side-effects are negligible (\SI{1}{ms}).
Thus, enabling malicious-server detection costs 1.4$\times$ in per-query latency: \SI{61}{ms} with verification vs.\ \SI{43}{ms} without.
The overhead is almost entirely the Freivalds checks themselves, whose trials stream the cluster matrix again.
Randomly verifying queries can reduce the verification cost significantly, at the risk that a forged answer on an unchecked query goes undetected.

\begin{takeaway}
The two privacy knobs cost different things: \sap{}'s $\beta$ trades off recall (0.93 to 0.70 for $\beta=0$ and $\beta=1$, respectively) for privacy and leaves latency unchanged.
\bntm{}'s verification has a nontrivial latency (1.4x) cost, almost entirely due to Freivalds checks.
\end{takeaway}

\begin{figure}[t]
  \centering
  \inputplot{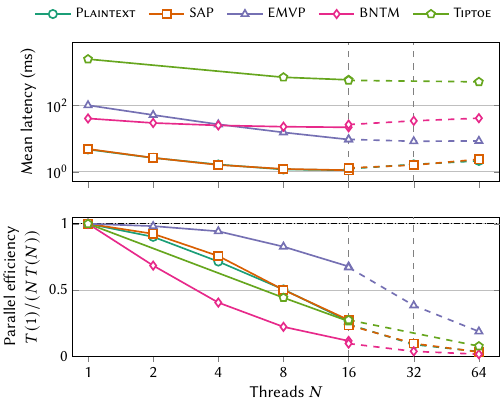}
  \caption{Performance of our schemes when increasing the number of threads, on \msmarcosmall. Top: mean latency vs thread count, log--log. Bottom: speed-up over the single-thread baseline. Solid lines: \ac{NUMA}-pinned to socket 0. Dashed lines: unpinned.}
  \Description{Six-line plot showing how each scheme's latency scales with thread count, with an inflection around N=16 where cross-socket coherence traffic becomes the bottleneck.}
  \label{fig:parallel-scaling}
\end{figure}

\subsection{Multi-threading}
\label{sec:eval-deploy}

\Cref{fig:parallel-scaling} shows how mean per-query latency (top) and parallel efficiency (bottom) changes as the CPU thread count $N$ increases from 1 to 64, thus measuring how multi-threading speeds up each scheme.
The sweep pins one thread per physical core of socket 0 up to 16 threads; larger counts run unpinned.
The top panel shows that latency falls up to 8--16 threads inside socket 0, but by less than the thread count grows.
\Cref{fig:parallel-scaling} (bottom) quantifies this as \emph{parallel efficiency} $T(1)/(N\,T(N))$, where $T(N)$ is the mean per-query latency at $N$ threads: 1 means ideal linear scaling, and adding threads without any latency reduction yields $1/N$.
At $N=16$ threads, parallel efficiency ranges from 0.68 (\emvp) down to 0.12 (\bntm).
The second socket adds little: only \emvp gets faster (14\,\% from $N=16$, pinned, to $N=32$, unpinned), the other schemes get slower because of cross-socket memory traffic.
Using all 64 logical threads (\ac{SMT}) increases latency for every scheme except \tiptoe, compared to 32 threads. %
With \ac{SMT}, two threads share one physical core's caches and memory ports; the score computation is already memory-bound, so the second thread adds contention without adding bandwidth.
The shape is the same for all schemes except \tiptoe: this is evidence that the bottleneck is the host's memory system, not the cryptographic primitive.

\begin{takeaway}
Shard per socket: only \emvp gains modestly from the second socket (14\,\%), and \ac{SMT} adds contention rather than throughput.
\end{takeaway}

\subsection{Index build time}
\label{sec:eval-index-build}
\begin{figure}[t]
  \centering
  \inputplot{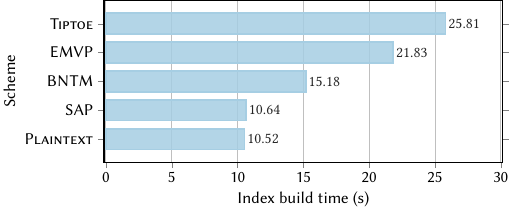}
  \caption{The time (in seconds) to build the index for each scheme from the \msmarcosmall corpus.
  }
  \Description{Horizontal bar chart with the scheme on the y-axis and cold build time in seconds. Bars are sorted by build time, all five schemes clustered between roughly 50 and 81 seconds.}
  \label{fig:build-time}
\end{figure}

Finally, \Cref{fig:build-time} reports the (one-time) compute time it takes per scheme by the client to construct the index from the \msmarcosmall corpus, \ie, running the $k$-means clustering and, for the schemes using encryption, encrypt the corpus vectors.
\plaintext builds the full index in \SI{10.5}{s}, which is essentially the $k$-means cost alone.
\sap has a marginal compute overhead because of the vector scaling and perturbation.
\bntm, \emvp and \tiptoe are slower (\SI{15.2}{s}, \SI{21.8}{s} and \SI{25.8}{s}, respectively), with all five schemes within a 2.5$\times$ band.
We attribute \tiptoe's index build time to the SimplePIR hint matmul computation.
All schemes share the \SI{10.5}{s} clustering cost.
The scheme-specific work on top of it ranges from negligible for \sap to \SI{11.3}{s} for \emvp's encryption and \SI{15.3}{s} for \tiptoe's hint computation, which match or exceed the clustering cost itself.

\begin{takeaway}
Index build takes tens of seconds at 100\,k: all schemes share the \SI{10.5}{s} $k$-means cost, and for \emvp and \tiptoe the encryption or hint computation costs about as much again.
\end{takeaway}
 
\section{Lessons Learned and Action Points}
\label{sec:discussion}
The key question for a deployment is which guarantee is worth which cost.

\paragraph{Action points.}
Our results translate into the following suggestions for practitioners working with vector search services:
\begin{itemize}
  \item If the server may see the queries and corpus, use \plaintext: every privacy scheme only adds cost.
  \item To hide the stored vectors at almost no performance cost, use \sap and pick the perturbation factor $\beta$ to match the recall the application requires; accept that the server can still learn the approximate corpus geometry.
  \item If corpus vectors and the query must remain hidden, use \emvp: budget a 4$\times$ throughput cost and 3$\times$ the memory bandwidth per query, and do not plan on a \ac{GPU}.
  \item If the server may also cheat, use \bntm and budget 1.4$\times$ its own latency for client-side response verification. %
  \item If the access pattern itself is sensitive, \tiptoe is the only option in this comparison but comes with 9$\times$ more per-query latency compared to \bntm. %
\end{itemize}

\paragraph{Implementation matters as much as protocol.}
A scheme's measured cost depends as much on the implementation as on the protocol: the \ac{CPU}-vs-\ac{GPU} asymmetry, for example, tracks memory bytes read per query rather than anything in the cryptographic construction.
We recommend that future evaluations hold the implementation substrate fixed and report protocol-level and implementation-level cost separately.

\section{Related Work}
\label{sec:related}

\paragraph{\ac{ANN} benchmarking.}
ANN-Benchmarks~\cite{aumuller:2020:annbenchmarks} is the
de-facto standard for comparing plaintext \ac{ANN} systems on a
common workload and metric, and the \textsc{BigANN}
challenge~\cite{simhadri:2022:bigann} extends it to billion-scale
corpora. Neither covers cryptographic
schemes. Our methodological contribution (\Cref{sec:methodology})
is to carry the common-harness idea to cryptographic schemes, so
that the cost of privacy can be read on the same axes as the cost
of recall.

\paragraph{Hardness assumptions.}
The four cryptographic schemes in this comparison rest on three
distinct hardness assumptions. \tiptoe{} is built on
\ac{LWE}~\cite{regev:2009:lwe}, the lattice assumption that
underlies most modern \ac{FHE} and \ac{PIR} constructions.
\bntm{} is built on \ac{LPN}~\cite{blum:1994:lpn}, a long-studied
assumption~\cite{pietrzak:2012:lpn} whose concrete hardness is
bounded by the BKW algorithm~\cite{blum:2003:bkw}. \emvp{} relies
on a more recent assumption (secret dual codes, introduced in the
original paper). The three occupy different points on the
conservativeness-performance trade-off, which the per-query costs
in \Cref{sec:evaluation} reflect.

\paragraph{Distance-comparison-preserving encryption.}
\sap{}~\cite{fuchsbauer:2021:dcpe} is the modern representative
of a longer line of work on order-revealing and order-preserving
encryption~\cite{boldyreva:2009:ope,popa:2013:mope,lewi:2016:ore},
which trades cryptographic indistinguishability for
the server's ability to operate directly on ciphertexts. \sap{}'s contribution is to
extend that idea to approximate distance comparisons over
real-valued vectors, as an \ac{IVF} index needs.

\paragraph{\ac{PIR}- and \ac{FHE}-based private retrieval.}
\tiptoe{}'s server-side computation builds on
SimplePIR~\cite{henzinger:2023:simplepir}, a low-overhead linearly
homomorphic \ac{PIR} scheme, and on \ac{BFV}~\cite{fan:2012:bfv}
for the per-query token. Earlier \ac{PIR} systems such as
XPIR~\cite{melchor:2016:xpir} and
SealPIR~\cite{angel:2018:sealpir} reach similar privacy at
substantially higher cost.

\paragraph{Encrypted matrix-vector products.}
\emvp{}~\cite{benhamouda:2025:emvp} and
\bntm{}~\cite{braverman:2026:bntm} sit in a small but active line
of work on encrypted linear algebra. A concurrent construction by
Vaikuntanathan and Zamir~\cite{vaikuntanathan:2025:cryptoeff}
explores recursive \acl{TDM} variants for the same problem.

\paragraph{Encrypted \ac{ANN} systems.}
\tiptoe{}~\cite{henzinger:2023:tiptoe} is the first end-to-end
private vector-search service. Adjacent work on private record retrieval, including
Splinter~\cite{wang:2017:splinter} and
PIR-PSI~\cite{demmler:2018:pirpsi}, achieves related goals on
structured key-value data rather than on dense vectors.

\paragraph{Trusted-hardware approaches.}
A different way to protect the vectors is to run the search
inside a trusted execution environment, such as Intel SGX. The
search then runs on plaintext inside the enclave at near-native
cost, but the trust shifts from cryptographic hardness to the
chip vendor's attestation: every server in the deployment must
support the chosen environment, and the enclave inherits the
vendor's side-channel track record.
The direction has been explored for encrypted
databases~\cite{zheng:2017:opaque} and oblivious key-value
stores~\cite{dauterman:2021:snoopy}, but we are not aware of a
published TEE-based vector-search system measured on axes
comparable to ours.

\section{Conclusion}
\label{sec:conclusion}

We presented a unified benchmark for privacy-preserving vector search that measured four cryptographic schemes (\sap{}, \emvp{}, \bntm{}, \tiptoe{}) against a \plaintext baseline on a single shared testbed: one \ac{IVF} index, one workload, one set of metric definitions, and the same hardware.
Two findings hold beyond the individual schemes.
First, wire bytes and memory bytes rank schemes differently: \bntm is cheap in network cost but expensive in memory access, and it is the memory requirements that predict the attainable speedup by using a \ac{GPU}.
Second, a scheme's cost is as much an implementation property as a protocol one.
These results debunk the claim that cryptographic privacy is unaffordable for vector search: it does not hold in our measurements on current hardware. 

\bibliographystyle{ACM-Reference-Format}

\appendix

\section{Appendix}
\label{sec:appendix}

\begin{figure}[H]
  \centering
  \inputplot{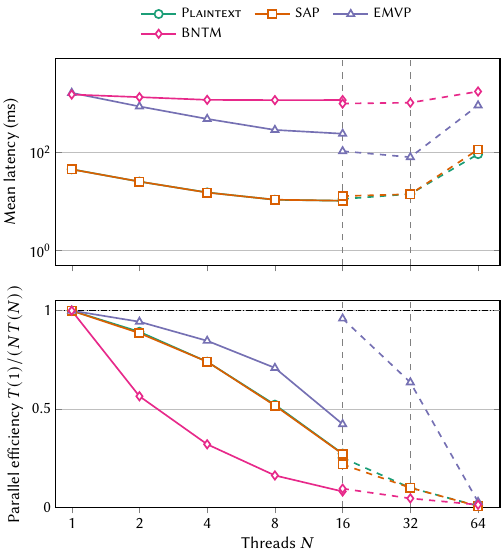}
  \caption{Parallel scaling on a dual-socket Xeon Gold 6426Y (\num{32} physical cores $+$ \ac{SMT}), \msmarcofull{}, all schemes except \tiptoe. Same shape as the main-body \plaintext-baseline figure but at the full corpus. \tiptoe is omitted from the 8.8\,M scaling sweep because no \tiptoe-\ac{GPU} build was measured and the \ac{CPU}-only path is out of scope for cross-substrate scaling at this size.}
  \Description{Parallel-scaling figure rebuilt from the 8.8 M MS MARCO sweep; same axes as the main-body parallel-scaling figure but with the larger corpus.}
  \label{fig:parallel-scaling-8M}
\end{figure}

\begin{figure}[H]
  \centering
  \inputplot{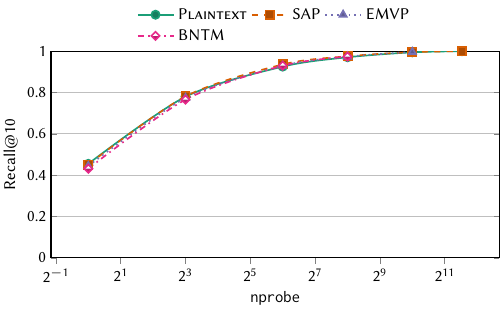}
  \caption{Recall@10 vs.\ \texttt{nprobe} for all schemes except \tiptoe{} on \msmarcofull. Reads off ``how aggressive does the probe set need to be to reach $X\%$ recall'' directly, instead of inferring from the latency / throughput Pareto in the main body. All four curves overlay tightly because every scheme shares the same plaintext-trained \ac{IVF} partition. \emph{Source:} \msmarcofull.}
  \Description{Four nearly-overlapping recall vs nprobe curves on a log-x axis: every scheme tracks the plaintext partition recall identically because they share the same cluster shape.}
  \label{fig:recall-nprobe}
\end{figure}

\begin{figure}[H]
  \centering
  \inputplot{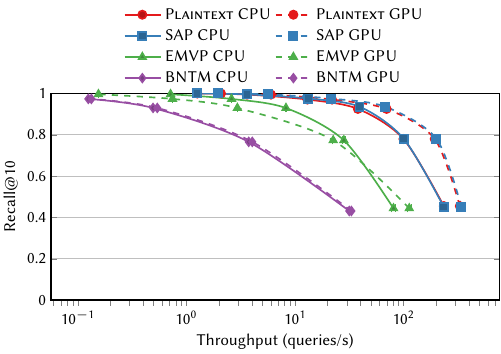}
  \caption{Recall@10 vs.\ sustained throughput, all schemes except \tiptoe, \ac{CPU} solid vs \ac{GPU} dashed of the same colour. \plaintext{} and \sap translate cleanly to the GPU (1.8--1.9$\times$ over the 64-thread CPU baseline); \emvp loses on the GPU because its encrypted matrix--vector kernel is memory-bandwidth-bound and, at 91\,GB, far exceeds device memory, so it streams over the PCIe bus rather than running resident; \bntm sits at single-digit qps on both sides. \emph{Source:} Xeon Gold 6426Y (64 cores) + RTX 5000 Ada, \msmarcofull.}
  \Description{Eight curves of recall versus throughput, paired CPU/GPU per scheme.}
  \label{fig:recall-throughput-cpu-vs-gpu}
\end{figure}

\begin{figure}[H]
  \centering
  \inputplot{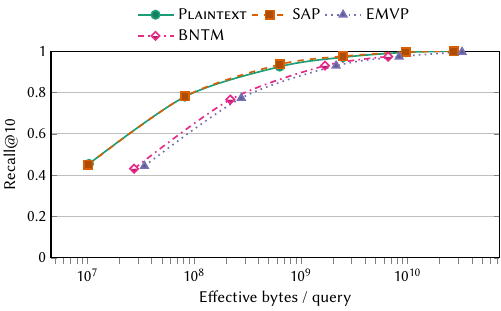}
  \caption{Recall@10 vs.\ \texttt{effective-bytes-per-query}, the analytical memory-bandwidth proxy that counts the bytes the server's matrix kernel streams locally per scoring call. Captures the per-query bandwidth premium the field-element matrices impose (2.7--3.4$\times$ over \plaintext\ for \emvp{} and \bntm{} at the same recall, even though their wire responses differ by orders of magnitude; \sap{} matches \plaintext{}) without requiring a per-scheme \ac{GPU} build. \emph{Source:} \msmarcofull.}
  \Description{Four curves of recall versus the analytical effective-bytes-per-query proxy on a log-x axis.}
  \label{fig:recall-effective-bytes}
\end{figure}

This appendix collects figures that supplement specific claims in
the main body but did not fit there.
\Cref{fig:parallel-scaling-8M} extends the parallel-scaling
reading to the full 8.8\,M corpus; \Cref{fig:recall-nprobe} is a
methodological sanity check on the shared \ac{IVF} partition; and
\Cref{fig:recall-throughput-cpu-vs-gpu,fig:recall-effective-bytes}
expand the \ac{CPU}-vs-\ac{GPU} reading from
\Cref{sec:eval-cpu-gpu}.

At the full 8.8\,M corpus the parallel-scaling shape replays as at
100\,k: the socket-boundary efficiency drop is the same for all
schemes except \tiptoe. \Cref{fig:parallel-scaling-8M} confirms
that the shared-memory bottleneck named in \Cref{sec:eval-deploy}
is corpus-invariant.

\Cref{fig:recall-nprobe} is the methodological sanity check that
lets the body attribute per-scheme cost differences to the
cryptographic primitive rather than to the index structure. All
schemes except \tiptoe{} overlay on the recall vs \texttt{nprobe}
curve because every scheme rides the same plaintext-trained
partition, so the probed-cluster set is bit-identical across
backends.

\Cref{fig:recall-throughput-cpu-vs-gpu} expands the
\Cref{tab:cpu-vs-gpu} summary into per-(scheme, substrate)
recall--throughput curves. \plaintext{} and \sap{} shift left of
their \ac{CPU} baselines on the \ac{GPU}; \emvp{} shifts right;
\bntm{} overlays itself on both substrates. The three regimes
named in \Cref{sec:eval-cpu-gpu} read off directly.

\Cref{tab:emvp-cpu-vs-gpu} isolates \emvp{}'s crossover: at
\texttt{nprobe}\,$=1$ a single small cluster fits on-device and the
\ac{GPU} wins, but as \texttt{nprobe} grows the per-query ciphertext
volume climbs, the matrix no longer fits in device memory, and the
\ac{CPU} pulls progressively ahead, reaching $4.6\times$ at
\texttt{nprobe}\,$=1024$. Recall is identical across substrates (the
kernels compute the same field matvec), so the gap is purely the
memory-bandwidth effect of \Cref{sec:eval-cpu-gpu}: the \ac{CPU} feeds
the matvec from \ac{DRAM} faster than the \ac{GPU} streams it over the
PCIe bus.

\begin{table}[H]
  \centering
  \footnotesize
  \caption{\emvp single-query throughput (qps, mean of three
  repetitions) on \ac{CPU} vs \ac{GPU} at \msmarcofull, by
  \texttt{nprobe}; recall@10 is identical on both substrates. The
  \ac{GPU} leads only at \texttt{nprobe}\,$=1$ (one small cluster,
  resident on-device); beyond that the encrypted matrix outgrows
  device memory and streams over the PCIe bus, so the \ac{CPU} wins by
  a margin that widens with the per-query byte volume (\texttt{CPU/GPU}
  $>1$ means \ac{CPU} faster). \emph{Source:} Xeon Gold 6426Y (64
  cores) + RTX 5000 Ada, \msmarcofull.}
  \label{tab:emvp-cpu-vs-gpu}
  \begin{tabular}{r r r r r}
    \toprule
    \texttt{nprobe} & recall@10 & CPU qps & GPU qps & CPU/GPU \\
    \midrule
    1    & 0.444 & 80.7 & 112.7 & 0.72 \\
    8    & 0.773 & 28.2 & 22.5  & 1.25 \\
    64   & 0.929 & 8.26 & 2.95  & 2.80 \\
    256  & 0.974 & 2.60 & 0.75  & 3.49 \\
    1024 & 0.996 & 0.72 & 0.16  & 4.61 \\
    \bottomrule
  \end{tabular}
\end{table}

\Cref{fig:recall-effective-bytes} plots recall against the
analytical \texttt{eff-bytes/q} proxy from \Cref{tab:byte-budget-8m}'s
mem-read column,
which counts the bytes the server's matrix kernel streams locally
per scoring call. The field-element schemes (\emvp{} and \bntm{})
sit 2.7--3.4$\times$ above \plaintext{} on the bandwidth axis
at the same recall, while \sap{}'s scale-and-perturb ciphertexts
stay f32-sized and match \plaintext{} byte-for-byte; this is the
in-memory cost that predicts the
\ac{GPU} asymmetry in \Cref{sec:eval-cpu-gpu} without requiring a
per-scheme \ac{GPU} build.

\subsection{Per-scheme parameters}
\label{sec:appendix-params}

\Cref{tab:scheme-params} consolidates the cryptographic parameters
of each scheme, in the source constructions' own notation, all at
the 128-bit security level used throughout (\Cref{sec:methodology}).
Every tuple targets the published $128$-bit ($\lambda{=}128$)
setting; the two choices that are ours rather than inherited are \emvp{}'s
zero-padding of the 768-dimensional embeddings up to the record
length $\ell_0 = 1024$ and \bntm{}'s $\lambda' = 3$ Freivalds trials
(the budget for $2^{-128}$ malicious-server detection).

\begin{table*}
  \centering
  \footnotesize
  \caption{Per-scheme parameters, in each construction's own notation, configured at a 128-bit level (\Cref{sec:methodology}); the basis differs by scheme, per the \emph{Security basis} column. For the cryptographic schemes (\emvp, \bntm, \tiptoe) it is the source paper's $\lambda{=}128$ computational-hardness analysis, which we adopt rather than re-derive. \sap{}/\ac{DCPE} is the exception: it carries a 128-bit \emph{symmetric key} (key secrecy), but has no computational-hardness assumption delivering a 128-bit bound --- its server-side privacy is the distance-distorting $\beta$ perturbation, not a $\lambda{=}128$ guarantee. \emvp{}'s listed tuple is our instantiation of the paper's $\lambda{=}128$ random-block regime (Table~1), not a tabulated row.}
  \label{tab:scheme-params}
  \begin{tabular}{l l l l}
    \toprule
    Scheme & Assumption & Parameters & Security basis \\
    \midrule
    \plaintext{} \(+\) \ac{IVF} & none                  & ---                                                                            & no cryptography \\
    \sap{} \(+\) \ac{IVF}       & symmetric (\ac{DCPE}) & $\beta \in \{0, 0.5\}$                                                          & key secrecy \\
    \emvp{} \(+\) \ac{IVF}      & secret dual codes     & $(n,k,s,b,\ell_0) = (1292,268,76,17,1024)$ ($768$-dim padded to $\ell_0$)       & $\lambda{=}128$ regime, Table~1~\cite{benhamouda:2025:emvp} \\
    \bntm{} \(+\) \ac{IVF}      & \ac{LPN}              & $(n,n_1,\delta,\varepsilon,\mu) = (1024,128,0.125,0.7,n^{-0.3})$, $\lambda' = 3$  & $\lambda{=}128$ target via the paper's heuristic (uncertified); $2^{-128}$ detection~\cite{braverman:2026:bntm} \\
    \tiptoe{}                   & \ac{LWE}              & $(n,\log_2 p,\sigma) = (2048,17,81\,920)$; \ac{BFV} at published params         & published params~\cite{henzinger:2023:tiptoe} \\
    \bottomrule
  \end{tabular}
\end{table*}

\subsection{Derivation of the communication budget}
\label{sec:appendix-comm-formulas}

\Cref{tab:byte-budget-100k} is derived from
closed-form expressions in the corpus geometry and the
per-scheme cryptographic parameters; this subsection states
those formulas and lists the parameter values used at each
corpus scale. The wire- and memory-byte budget is deterministic
once these are fixed, which is what lets us project \tiptoe{}'s
8.8\,M row without an end-to-end run at that scale.
The 8.8\,M corpus needs \texttt{nprobe} = 64 to reach the same recall.

\paragraph{Parameters.}
\Cref{tab:comm-params} lists the symbols. The corpus geometry
($c$, $m$, $m_{\max}$) is determined by the
\acl{IVF} partitioning of the corpus; the scheme parameters are
fixed by each construction's security level (a 128-bit level across
all schemes --- cryptographic-hardness $\lambda{=}128$ for \emvp,
\bntm, and \tiptoe; a 128-bit symmetric key for \sap, whose privacy
is the $\beta$ perturbation rather than a hardness bound; see
\Cref{sec:bg-schemes} and \Cref{tab:scheme-params}). The \tiptoe{} parameters
$n_{\textrm{LWE}}$, the \ac{BFV} polynomial degree
$\rho_{\textrm{BFV}}$, the per-ciphertext byte cost
$b_{\textrm{BFV}}$, and the limb count $L$ come directly from
the SimplePIR\,/\,\ac{BFV} security configuration used in the
implementation~\cite{henzinger:2023:tiptoe,henzinger:2023:simplepir,fan:2012:bfv}.

\begin{table}[H]
  \centering
  \caption{Communication-budget parameters used in
  \Cref{tab:byte-budget-100k}.}
  \label{tab:comm-params}
  \resizebox{\columnwidth}{!}{%
  \begin{tabular}{l l r r}
    \toprule
    Symbol & Meaning & 100\,k & 8.8\,M \\
    \midrule
    $N$                       & corpus size                                & 100\,000        & 8\,800\,000     \\
    $c$                       & cluster count ($\lceil \sqrt{N} \rceil$)   & 317             & 2\,967          \\
    $m$                       & mean cluster size ($N / c$)                & 315             & 2\,966          \\
    $m_{\max}$                & largest cluster (Tiptoe padding target)    & 977             & 9\,186$^{\ddagger}$ \\
    \texttt{nprobe}           & probed clusters per query at recall@10$\approx 0.9$ & 32     & 64              \\
    $k$                       & top-$k$ results returned                   & 10              & 10              \\
    $d$                       & embedding dimension                        & 768             & 768             \\
    $F$                       & field byte width (u64) for crypto schemes  & 8               & 8               \\
    $n_{\textrm{EMVP}}$       & \emvp{} encoded-row width                  & 1292            & 1292            \\
    $s_{\textrm{EMVP}}$       & \emvp{} block count per cluster            & 76              & 76              \\
    $n_{\textrm{BNTM}}$       & \bntm{} encoded-row width                  & 1024            & 1024            \\
    $n_{\textrm{LWE}}$        & Tiptoe \ac{LWE} dimension                  & 2048            & 2048            \\
    $\rho_{\textrm{BFV}}$     & Tiptoe \ac{BFV} polynomial degree          & 4096            & 4096            \\
    $b_{\textrm{BFV}}$        & per-\ac{BFV}-ciphertext byte cost          & 20\,480         & 20\,480         \\
    $L$                       & Tiptoe \ac{BFV} limb count                 & 8               & 8               \\
    \bottomrule
  \end{tabular}}
  \par\vspace{2pt}\footnotesize $^{\ddagger}$\,The 8.8\,M value of $m_{\max}$ is projected from the 100\,k max-to-mean ratio ($977 / 315 \approx 3.10$) applied to the 8.8\,M mean cluster size; we have not partitioned the 8.8\,M corpus through Tiptoe's encoder. The assumption is that k-means cluster-size dispersion is roughly scale-invariant for the same data distribution.
\end{table}

\paragraph{Per-scheme formulas.}
\Cref{tab:comm-formulas} states the closed-form expression
behind each cell in \Cref{tab:byte-budget-100k}.
Three observations:
(i) the online \emph{query} ($\uparrow$) is a function of the
scheme's input encoding only, not of corpus size, except for
\tiptoe{} where the encrypted \ac{LWE} vector has one slot per
$(\textrm{cluster}, \textrm{dimension})$ pair and therefore
grows with $c$;
(ii) the per-query \emph{response} ($\downarrow$) of the
encrypted-score schemes scales with \texttt{nprobe} and with the
per-cluster row count $m$ (or $m_{\max}$ for \tiptoe{}, since the
SimplePIR answer covers the padded cluster); \plaintext{} and
\sap{} score in the clear on the server and return only the
global top-$k$ (a \SI{4}{B} identifier and a \SI{4}{B} score per
hit, \SI{80}{B} at $k = 10$), independent of \texttt{nprobe} and
corpus size;
(iii) the one-time \emph{setup} for the \ac{IVF} schemes scales
linearly with the corpus ($c \cdot m \cdot n \cdot F$), so the
8.8\,M block of \Cref{tab:byte-budget-8m} sits
88$\times$ above the 100\,k block.

\begin{table*}
  \centering
  \footnotesize
  \caption{Closed-form byte formulas behind each column of \Cref{tab:byte-budget-100k}. Symbols are defined in \Cref{tab:comm-params}. Let $p = \texttt{nprobe}$.}
  \label{tab:comm-formulas}
  \setlength{\tabcolsep}{4pt}
  \begin{tabular}{l l l l l l l}
    \toprule
    Scheme & query $\uparrow$ & response $\downarrow$ & setup & offline $\uparrow$ & offline $\downarrow$ & \texttt{eff-bytes/q} \\
    \midrule
    \plaintext{} \(+\) \ac{IVF}  & $4 d$                          & $k \cdot (4 + 4)$                      & ---                                       & ---                                  & ---                                                                          & $p \cdot m \cdot 4 d$     \\
    \sap{} \(+\) \ac{IVF}        & $F d$                          & $k \cdot (4 + 4)$                      & $c \cdot m \cdot F d$                     & ---                                  & ---                                                                          & $p \cdot m \cdot 4 d$     \\
    \emvp{} \(+\) \ac{IVF}       & $F n_{\textrm{EMVP}}$          & $p \cdot s_{\textrm{EMVP}} \cdot m \cdot F$  & $c \cdot m \cdot n_{\textrm{EMVP}} \cdot F$ & ---                                  & ---                                                                          & $p \cdot m \cdot n_{\textrm{EMVP}} \cdot F$ \\
    \bntm{} \(+\) \ac{IVF}       & $F n_{\textrm{BNTM}}$          & $p \cdot m \cdot F$                    & $c \cdot m \cdot n_{\textrm{BNTM}} \cdot F$ & ---                                  & ---                                                                          & $p \cdot m \cdot n_{\textrm{BNTM}} \cdot F$ \\
    \tiptoe{}                    & $c \cdot d \cdot F$            & $m_{\max} \cdot F$                     & $m_{\max} \cdot n_{\textrm{LWE}} \cdot F$ & $n_{\textrm{LWE}} \cdot b_{\textrm{BFV}}$ & $\lceil m_{\max} / \rho_{\textrm{BFV}} \rceil \cdot L \cdot b_{\textrm{BFV}}$ & $m_{\max} \cdot n_{\textrm{LWE}} \cdot F$ \\
    \bottomrule
  \end{tabular}
\end{table*}

\paragraph{Extrapolation to 8.8\,M.}
The 8.8\,M block of \Cref{tab:byte-budget-8m} applies
\Cref{tab:comm-formulas} at the 8.8\,M cluster geometry. All rows except \tiptoe's are end-to-end
measurements from runs at this scale; \tiptoe{}$^\dagger$ is
computed from the formulas, since the \ac{LWE} matvec was not
run end-to-end at 8.8\,M (the \ac{LWE} element width pushes the
end-to-end run out of scope for our hardware budget). The 100\,k
agreement between formulas and measurements is what justifies
trusting the \tiptoe{} projection here.

\paragraph{Worked examples.} Substituting
\Cref{tab:comm-params} into \Cref{tab:comm-formulas} reproduces
each cell of \Cref{tab:byte-budget-100k,tab:byte-budget-8m}. Two
representative cases:
\begin{itemize}
  \item \bntm{} setup at 8.8\,M: $c \cdot m \cdot
    n_{\textrm{BNTM}} \cdot F = 2967 \times 2966 \times 1024
    \times 8 \approx \SI{72.1}{GB}$.
  \item \tiptoe{} \ac{BFV} offline $\uparrow$ at either scale:
    $n_{\textrm{LWE}} \cdot b_{\textrm{BFV}} = 2048 \times 20\,480
    = \SI{42}{MB}$. This term has no $N$ dependence — it is the
    cost of transmitting one \ac{BFV} ciphertext per \ac{LWE}
    secret-vector slot, and the \ac{LWE} dimension is fixed by
    the security level.
\end{itemize}

\paragraph{Analytical vs.\ realised values.} The formulas use
the partition-wide mean cluster size $m = N / c$. The per-query
realised cost reported by the harness uses the empirical mean
over the probed cluster set, which at 8.8\,M tracks
10--15\,\% higher than $N / c$ because (i) k-means does
not produce perfectly balanced clusters at this scale and (ii)
\ac{IVF} routing biases the probe set toward denser clusters.
The 8.8\,M block of \Cref{tab:byte-budget-8m} reports the
realised values for all rows except \tiptoe's; the 100\,k block
agrees with the formulas within rounding because the cluster-size variance is
small at the 100\,k validation scale. Setup and \tiptoe{}'s
offline-phase costs are independent of routing and match the
formulas exactly at both scales.
\sap{}'s setup entries are computed from the formula
($c \cdot m \cdot F d$) rather than measured.
 
\end{document}